\documentclass[preprint2]{aastex}    

\usepackage{graphicx}
\usepackage{subfigure}
\usepackage{amsmath,epsfig,tabularx,amssymb,float}
\usepackage[latin1]{inputenc}
\usepackage{times}
\usepackage{paralist}
\usepackage{latexsym}   
\usepackage{color}
\usepackage{theorem}
\usepackage{textcomp}

\DeclareMathAlphabet{\mathfat}{U}{bbold}{m}{n}

\newtheorem{theorem}{Theorem}
\newtheorem{lemma}[theorem]{Lemma}

\newtheorem{prop}{Proposition}
\newtheorem{claim}{Claim}

\newtheorem{definition}{Definition}
\newtheorem{question}{Question}
\newtheorem{coro}{Corollary}

\newcommand{\beq}{\begin{equation}}
\newcommand{\enq}{\end{equation}}

\newcommand{\bea}{\begin{array}}
\newcommand{\ena}{\end{array}}
\newcommand{\bds}{\begin {itemize}}
\newcommand{\eds}{\end {itemize}}
\newcommand{\bdf}{\begin{definition}}
\newcommand{\blm}{\begin{lemma}}
\newcommand{\edf}{\end{definition}}
\newcommand{\elm}{\end{lemma}}
\newcommand{\bthm}{\begin{theorem}}
\newcommand{\ethm}{\end{theorem}}
\newcommand{\bprp}{\begin{prop}}
\newcommand{\eprp}{\end{prop}}
\newcommand{\bcl}{\begin{claim}}
\newcommand{\ecl}{\end{claim}}
\newcommand{\bcr}{\begin{coro}}
\newcommand{\ecr}{\end{coro}}
\newcommand{\bquest}{\begin{question}}
\newcommand{\equest}{\end{question}}

\newcommand{\larrow}{{\larrow}}

\newcommand{\argmin}{\ensuremath{\mathrm{arg}\min}}
\newcommand{\argmax}{\ensuremath{\mathrm{arg}\max}}

\newcommand{\tr}{\ensuremath{\mathrm{tr}}}

\newcommand{\va}{{\ensuremath{{\mathbf{a}}}}}

\newcommand{\vf}{{\ensuremath{{\mathbf{f}}}}}

\newcommand{\vn}{{\ensuremath{{\mathbf{n}}}}}

\newcommand{\vu}{{\ensuremath{{\mathbf{u}}}}}

\newcommand{\vw}{{\ensuremath{{\mathbf{w}}}}}

\newcommand{\vx}{{\ensuremath{{\mathbf{x}}}}}

\newcommand{\mA}{{\ensuremath{\mathbf{A}}}}

\newcommand{\mC}{{\ensuremath{\mathbf{C}}}}

\newcommand{\mI}{{\ensuremath{\mathbf{I}}}}

\newcommand{\mP}{{\ensuremath{\mathbf{P}}}}

\newcommand{\mQ}{{\ensuremath{\mathbf{Q}}}}

\newcommand{\mR}{{\ensuremath{\mathbf{R}}}}

\newcommand{\mU}{{\ensuremath{\mathbf{U}}}}

\newcommand{\mV}{{\ensuremath{\mathbf{V}}}}

\newcommand{\mX}{{\ensuremath{\mathbf{X}}}}
\newcommand{\mY}{{\ensuremath{\mathbf{Y}}}}
\newcommand{\mZ}{{\ensuremath{\mathbf{Z}}}}

\usepackage{latexsym}

\def\IC{\mathbb C}
\def\IN{\mathbb N}
\def\IZ{\mathbb Z}
\def\IR{\mathbb R}

\def\shat{^{\mathchoice{}{}%
 {\,\,\smash{\hbox{\lower4pt\hbox{$\widehat{\null}$}}}}%
 {\,\smash{\hbox{\lower3pt\hbox{$\hat{\null}$}}}}}}

\def\bSigma{{
      \ooalign{
      \smash{\hskip.4pt\raise.4pt\hbox{$\Sigma$}}\vphantom{}\crcr
      \smash{\hskip.7pt\raise.6pt\hbox{$\Sigma$}}\vphantom{}\crcr
      \smash{\hbox{$\Sigma$}}\vphantom{$\Sigma$}}
      \vphantom{\hbox{$\Sigma$}}
      }}
\def\bTheta{{
      \ooalign{
      \smash{\hskip.5pt\raise.5pt\hbox{$\Theta$}}\vphantom{}\crcr
      \smash{\hskip.0pt\raise.1pt\hbox{$\Theta$}}\vphantom{}\crcr
      \smash{\hbox{$\Theta$}}\vphantom{$\Theta$}}
      \vphantom{\hbox{$\Theta$}}
      }}
\def\bDelta{{
      \ooalign{
      \smash{\hskip.4pt\raise.4pt\hbox{$\Delta$}}\vphantom{}\crcr
      \smash{\hskip.7pt\raise.6pt\hbox{$\Delta$}}\vphantom{}\crcr
      \smash{\hbox{$\Delta$}}\vphantom{$\Delta$}}
      \vphantom{\hbox{$\Delta$}}
      }}

\makeatletter

\def\bordermatrix#1{\begingroup \m@th
  \@tempdima 8.75\p@
  \setbox\z@\vbox{%
    \def\cr{\crcr\noalign{\kern2\p@\global\let\cr\endline}}%
    \ialign{$##$\hfil\kern2\p@\kern\@tempdima&\thinspace\hfil$##$\hfil
      &&\quad\hfil$##$\hfil\crcr
      \omit\strut\hfil\crcr\noalign{\kern-\baselineskip}%
      #1\crcr\omit\strut\cr}}%
  \setbox\tw@\vbox{\unvcopy\z@\global\setbox\@ne\lastbox}%
  \setbox\tw@\hbox{\unhbox\@ne\unskip\global\setbox\@ne\lastbox}%
  \setbox\tw@\hbox{$\kern\wd\@ne\kern-\@tempdima\left[\kern-\wd\@ne
    \global\setbox\@ne\vbox{\box\@ne\kern2\p@}%
    \vcenter{\kern-\ht\@ne\unvbox\z@\kern-\baselineskip}\,\right]$}%
  \null\;\vbox{\kern\ht\@ne\box\tw@}\endgroup}
\makeatother

\newcommand{\DL}{\begin{dashlist}}
\newcommand{\DLE}{\end{dashlist}}

\makeatletter
\def\argmin{\mathop{\operator@font arg\,min}}
\def\argmax{\mathop{\operator@font arg\,max}}
\makeatother

\renewcommand{\tr}{H}

\newcommand{\beqa}{\begin{eqnarray}}
\newcommand{\enqa}{\end{eqnarray}}

\begin{document}

\title{
Phased Array Feed Calibration, Beamforming and Imaging }

\shorttitle{Beamforming for Array Feeds}
\shortauthors{Landon et al.}

\author{Jonathan Landon, Michael Elmer, Jacob Waldron, David Jones, Alan Stemmons, \\ Brian D. Jeffs, Karl F. Warnick}
\affil{Department of Electrical and Computer Engineering \\
  Brigham Young University \\
  459 Clyde Building,
  Provo, UT  84602, USA}

\author{J. Richard Fisher}
\affil{National Radio Astronomy Observatory\altaffilmark{1} \\
   Charlottesville, VA, USA}

\author{Roger D. Norrod}
\affil{National Radio Astronomy Observatory\altaffilmark{1} \\
  Green Bank, WV, USA}

\altaffiltext{1}{The NRAO is operated for the National Science Foundation (NSF) by Associated Universities, Inc.\ (AUI) under a cooperative agreement.}

\begin{abstract}
Phased array feeds (PAFs) for reflector antennas offer the potential for increased reflector field of view and faster survey speeds.  To address some of the development challenges that remain for scientifically useful PAFs, including calibration and beamforming algorithms, sensitivity optimization, and demonstration of wide field of view imaging, we report experimental results from a 19 element room temperature L-band PAF mounted on the Green Bank 20-Meter Telescope.  Formed beams achieved an aperture efficiency of 69\% and system noise temperature of 66 K.  Radio camera images of several sky regions are presented.  We investigate the noise performance and sensitivity of the system as a function of elevation angle with statistically optimal beamforming and demonstrate cancelation of radio frequency interference sources with adaptive spatial filtering.
\vspace{.2in}
\end{abstract}

\section{Introduction}
\label{sec:intro}

The generation of radio astronomy instruments now coming on line will fill much of the observational parameter space---frequency coverage, instantaneous bandwidth, system noise temperature, angular resolution,
and time and frequency resolution.  The two remaining open-ended parameters where fundamentally new science will be explored are collecting area and field of view.  More collecting area will allow us to observe known phenomena much deeper in the universe, and greater fields of view open the possibility of finding new phenomena, such as transient radio sources and rare types of pulsars, by making large-area sky surveys much more efficient.  Several major instrument development projects are underway worldwide to develop radio cameras, variously referred to as active, phased, beamforming, or smart arrays to distinguish them from the more conventional independent-pixel feed-horn arrays which sample less than 1/16th of the available sky area within the array's field-of-view.  These instruments will utilize focal plane phased array feeds (PAF) which can electronically synthesize multiple, simultaneous far field beams for complete coverage of the field of view without loss of sensitivity in each beam. PAFs are in development or in planning stages for single dish instruments including the Green Bank Telescope (GBT) as well as the Westerbork Synthesis Radio Telescope (WSRT) \citep{apertif_08} and the Australian Square Kilometer Array (ASKAP) Pathfinder synthesis imaging array \citep{askap_eucap_07}.

As the cost of signal processing for beam-forming and the correlation of multiple beams in aperture synthesis arrays allows, wider fields of view for SKA and its predecessors will open up new science, such as surveys of neutral hydrogen at high redshifts to study the evolutionary history of the universe.  First science that will be enabled by PAFs with modest signal processing bandwidths include observations
of the trajectories and abundance of high-velocity HI clouds interacting with the disk of the Milky Way \citep{wakker_97,putman_06,lockman_08}, surveys of gas clouds outside other galaxies covering large areas \citep{thilker_04,braun_04,grossi_08}, and studies of the kinematics of extended HI clouds contained by galactic groups \citep{borthakur_08,braun_04,chynoweth_08}.  These types of observations are time consuming with single-pixel instruments and could be conducted more rapidly and to greater depth using a PAF with wide field of view.

Demand for use of the GBT to study the inner parts of the Milky Way disk is extremely high.  Wider fields of view would enable additional studies of the galactic center and objects such as the physical conditions and structure of the recently discovered ``far'' counterpart to the 3-kpc arm \citep{dame_08}.  A PAF on the GBT would be of immediate use in analyzing the complex chemistry and relationship to dust evolution of the diffuse interstellar medium using 18cm emission from OH, and in understanding the medium to minimize the effect of foreground matter on cosmological investigations \citep{black_77,gillmon_06,liszt_96,leach_08}.

As compared to single-pixel instruments, PAF beam formation requires a substantial amount of signal processing for array calibration, operational beamforming, and image formation.  Calibration and beamforming algorithms and performance criteria for array feeds were surveyed by \cite{Jeffs08b}.  Radio frequency interference (RFI) mitigation using adaptive nulling algorithms was demonstrated by \cite{array_feed_exp_07}.  Other recent results demonstrating feasibility of PAF-based interference mitigation
in astronomical applications include \citep{Jeffs07,Jeffs07icassp,Jeffs08b,Landon08,Landon08b,Warnick07antProp}.

In this paper, we present results from an L-band prototype PAF on the Green Bank 20-Meter Telescope.  Early radio camera images obtained with the PAF were reported by \cite{paf_warnick_ursi08}.  A more detailed study of performance metrics including sensitivity and system noise are given here.  A major concern with PAF development has been optimizing the interface between the array and front end receiver electronics \citep{opt_match,array_ivashina_08}.  Due to mutual coupling between array antenna elements, standard impedance matching techniques used for single-port antennas must be extended to multi-port systems using the theory of active impedances \citep{active_woestenburg_tr_05,mc_warnick_08}.  The performance improvement that could be realized using these methods are explored.

PAFs offer the potential for optimizing beam patterns to improve sensitivity as the noise environment changes or to maximize system performance for a given type of observation.  We study the dependence of sensitivity and efficiency of formed beams on the reflector tipping angle as sky noise increases near the horizon.  Results of RFI mitigation experiments are also shown, in order to demonstrate the potential for operational use of RFI nulling algorithms in observations. These results are strong evidence for the feasibility of a high sensitivity, wide field of view PAF.

\section{Experiment Description}
\label{sec:ground_shield}

\begin{figure}[t]
\begin{center}
\epsfig{file=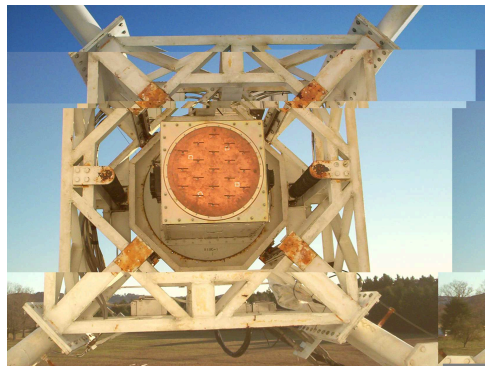,width=2.5in}
\end{center}
\caption{
19 element single polarized PAF and front end box containing analog downconverters and receiver front ends, mounted on on Green Bank 20-Meter Telescope (October, 2007).
\label{fig:20meter}}
\end{figure}

\begin{figure}[t]
\begin{center}
\includegraphics[width=2.5in]{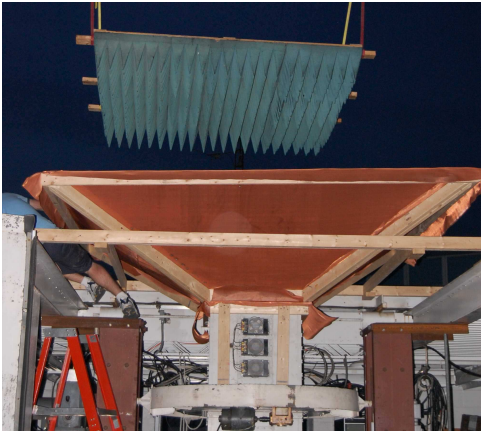}
\end{center}
\caption{
Ground shield and PAF in sky noise measurement facility with absorber being lowered over the array for $T_{\rm hot}$ measurement measurement (July, 2008).
\label{fig:SkyNoiseFacility}}
\end{figure}

The characterization techniques, calibration methods, and beamforming algorithms used in the reported work were developed with a prototype 19 element array (Fig.\ \ref{fig:20meter}) of single polarized, thickened dipoles backed by a ground plane, with a 20 channel non-cryocooled receiver and data acquisition system for real time sampling and streaming to disk.  The array was initially characterized using the BYU ``Very Small Array'' 3-meter prototype platform and the NRAO antenna range in Green Bank, WV.  In October 2007 and July 2008 the prototype PAF was mounted on the Green Bank 20-Meter Telescope to measure aperture and spillover efficiency, demonstrate multiple beam formation, and test RFI mitigation algorithms.

\subsection{Array Feed and Data Acquisition}

\begin{figure}[t]
\begin{center}
\epsfig{file=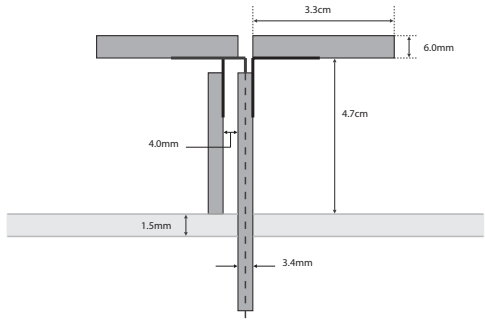,width=2.5in}
\end{center}
\caption{
Dipole element diagram.
\label{fig:dipole}}
\end{figure}

The PAF was located at the focus of the Green Bank 20-Meter 0.43 $f/D$ reflector.  Array elements were balun-fed half-wave dipoles tuned to 1600MHz $\lambda/4$ above a ground plane.  The dipoles achieved a 570 MHz impedance bandwidth (-10 dB reflection coefficient) or a 35\% bandwidth relative to the center frequency, which is modest but adequate for initial tests.  The elements are arranged in two concentric hexagonal rings around a center dipole.  The hexagonal grid permits 0.6$\lambda$ spacing between dipoles, which is slightly farther apart than the typical $\lambda$/2 for a rectangular grid.  This increased inter-element spacing reduced mutual coupling without undersampling or producing grating lobes.  A diagram of the dipole element is shown in Figure \ref{fig:dipole}.  Additional details were given by \cite{array_feed_exp_07}.

The dipoles were connected through the array ground plane to uncooled LNAs (Ciao Wireless, Inc., Camarillo, CA) with 33 K noise temperature and 41 dB gain at 1600 MHz.  The LNAs were measured by L.\ Belostotski (U.\ Calgary) to have the following noise parameters at 1600 MHz: $T_{\rm min} = 33\,{\rm K}$, $R_{\rm n} = 3.4\,\Omega$, and $\Gamma_{\rm opt} = 0.07 \angle 90^{\circ}$.

Receivers in the front-end box at the dish focus performed a two-stage downconversion to a final IF centered at 2.8125 MHz with a 3 dB bandwidth of 425 kHz.  IF signals were fed by coaxial cable into the pedestal room of the telescope to a data acquisition system in a shielded rack.  IF voltages were synchronously baseband-subsampled using 12-bit quantization for high dynamic range at a sample rate of 1.25 Msamples/sec per channel for 20 channels, using five commercial data acquisition cards with four channels per card.  A single network server class PC housed all acquisition cards and a high speed array of hard drives for real-time data streaming to disk. Beams were formed in post-processing on the raw voltage samples.

\subsection{Isotropic Noise Measurement Facility}

Measurements of the array isotropic noise response can be used to determine the aperture efficiency and system noise temperature associated with formed beams \citep{effic_warnick_08}.  The array was mounted in a facility with a retractable roof to produce a noise field that is approximately isotropic at the temperature of cold sky over the array's significant response directions.  A copper screen connected to the array ground plane at a slope of $30^{\circ}$ provided partial shielding of the array from noise radiated by ground, horizon, and nearby buildings.  This provided the cold load for a $Y$-factor array isotropic noise temperature measurement.  For the hot load, a 2.4m $\times$ 2.4m sheet of RF absorber was lowered over the copper screen (Fig.\ \ref{fig:SkyNoiseFacility}).  The computed array output voltage correlation matrices for the two configurations were used to characterize the system noise for the results given in Section \ref{sec:hotCold}.

\section{PAF Beamforming}

\begin{figure}[t]
\begin{center}
\epsfig{file=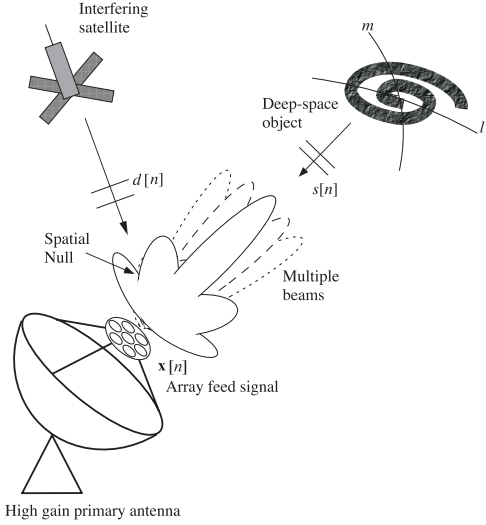,width=2.8in}
\end{center}
\caption{
The primary advantage of FPA telescopes is increased field of view provided by multiple, simultaneously formed beams.
Spatial cancelation of interfering signals is also possible.
\label{fig:multibeam}}
\end{figure}

In this section we discuss principles and methodologies for practical beamforming with an astronomical phased array feed instrument. Figure \ref{fig:multibeam} illustrates how digital beamforming techniques enable a PAF to form multiple simultaneous beams to increase the instantaneous field of view for the telescope.

\subsection{Signal Model}
\label{sec:SigMod}

We assume that the array output signals are processed in narrow subbands such that $ B \ll D/c$, where $B$ is bandwidth, $D$ is the PAF diameter, and $c$ is the speed of light.  After downconversion and sampling, the $M$ element PAF  produces a length $M \times 1$ complex baseband data vector at time sample $n$:
\begin{eqnarray}
\vx[n] & = & \va s[n]  + \vn[n]
\end{eqnarray}
and where $s[n]$ is the signal of interest (SOI) and $\vn[n]$ is the total system noise vector seen at the array. In the presence of interference, $\vn[n]$ will also include components due to a man-made detrimental signal $d[n]$ as illustrated in Fig.\ \ref{fig:multibeam}.  Vector $\va$ is the normalized array response
to a unit amplitude point source in the far field direction corresponding to $s[n]$.

Assuming zero mean wide-sense stationary signals and statistical independence between distinct signal and noise sources the array covariance is
\begin{eqnarray}
\mR & = & {\rm E}\{ \vx [n] \vx^H [n] \} \; = \; \mR_{\rm s} + \mR_{\rm n}  \\
\mR_{\rm s} & = & \sigma_s^2 \va \va^H \nonumber \\
 \mR_{\rm n} & =  &  \mR_{\rm rec} +  \mR_{\rm sp} +  \mR_{\rm sky} +  \mR_{\rm loss} + \mR_{\rm int}
\label{eq:noiseCovars}
\end{eqnarray}
where $\sigma_s^2$ is SOI signal power (variance), and $\mR_{\rm rec}$,  $\mR_{\rm sp}$,  $\mR_{\rm sky}$, $\mR_{\rm loss}$ and $\mR_{\rm int}$ are receiver, spillover, sky, antenna resistive loss, and interference noise covariance matrices, respectively.
$\mR_{\rm sky}$ refers to non-SOI noise sources in the beam main lobe, including atmospheric, cosmic microwave background (CMB) and galactic background (GB) radiation.
Models for each of these terms, including their spatial structure across the array and effective noise temperatures seen at the beamformer output, will be introduced in Section \ref{sec:noiseModels}.
Except in Section \ref{sec:rfi} where we consider interference canceling, we will assume $\mR_{\rm int} = \boldsymbol{0}$.

$\mR$ is estimated for the $j$th short term integration
(STI) window with sample covariance matrix
\begin{eqnarray}
\hat{\mR}^{(j)} & = & \frac{1}{L} \sum_{n=jL}^{(j+1)L-1} \vx [n] \vx^H [n]   \; = \;  \frac{1}{L} \mX_j  \mX_j^H   \label{eqn:R_x_hat} \\
\mX_j & = & \left[ \rule{0mm}{2.0ex} \vx[jL], \vx[jL+1] , \cdots , \vx[(j+1)L-1] \right] .
 \nonumber
\end{eqnarray}
where $\mX_j$ is the $L$ sample long STI data window.
The required length $L$ depends on the signal scenario and desired operational mode.
If adaptive cancelation is used to suppress moving interference as discussed in Section \ref{sec:rfi}, beamformer weights (computed from $\hat{\mR}^{(j)}$)  must be updated rapidly with $L$ short enough that $\mR_{\rm int}$ does not change over the STI window.
A slower update rate, and thus longer $L$, is used to adapt to spillover and sky noise due to pointing changes, as discussed in Section \ref{sec:tipping}.
Finally, in stable signal conditions, $L$ can  be very long (on the order of minutes or hours) to reduce sample estimation error and yield accurate calibration or high sensitivity beamformer solutions.
In these cases superscript $(j)$ will be dropped for simplicity when this does not lead to an ambiguous interpretation.

\subsection{Calibration}
\label{sec:Cal}

Since multiple simultaneous beams are formed with a PAF as shown in Fig.~\ref{fig:multibeam}, a calibration for the signal array response vector $\va_i$ must be performed for each direction, $\Omega_i$, corresponding to each formed beam's boresight direction, and any additional directions where point constraints in the beam pattern response will be placed.  Periodic re-calibration may be necessary due to strict beam pattern stability requirements, to correct for differential electronic phase and gain drift, and to characterize changes in receiver noise temperatures.

We proposed a calibration procedure in \citep{Jeffs08b}, which is improved here with a lower error, noise whitening approach in step \ref{calStep3}.

\noindent {\bf \small Calibration Procedure:}
\begin{enumerate}

\item \label{calStep1} {\em Noise covariance}, $\breve{\mR}_{\rm n}$:  Steer the dish to a relatively empty patch of sky so $\vx [n] \approx \vn [n]$, and collect a long term (large $L$, e.g.\ 10 minutes) sample covariance estimate for the noise field $\breve{\mR}_{\rm n}$ using (\ref{eqn:R_x_hat}).

\item  \label{calStep2} {\em Signal-plus-noise covariances}, $\breve{\mR}_{i}$:  While tracking the brightest available calibration point source,
steer the dish to calibration angle $\Omega_i$ (relative to this source).
The observed signal model is $\vx [n] = \va_{i} s[n] + \vn [n]$, where $\va_{i}$ is the desired calibration vector at direction $\Omega_i$.
Calculate $\breve{\mR}_{i}$ using (\ref{eqn:R_x_hat}) and the same $L$ as in step \ref{calStep1}.

\item \label{calStep3} {\em Array response,} $\breve{\va}_{i}$:  Compute $\breve{\va}_{i} = \breve{\mR}_{\rm n} \vu_{\rm max}$, where $\vu_{\rm max}$ is the dominant solution to the generalized eigenequation $\breve{\mR}_i \vu_{\rm max} = \lambda_{\rm max} \breve{\mR}_{\rm n} \vu_{\rm max}$.  This noise-whitening method produces a lower variance estimate with less bias arising from correlated noise in $\mR_{\rm n}$ (caused by inter-element mutual coupling) than does the method of \cite{Jeffs08b}.

\item \label{calStep4} {\em SOI covariance,}  $\breve{\mR}_{{\rm s},i}$:  Compute $\breve{\mR}_{{\rm s},i} = \lambda_{\rm max} \breve{\va}_{i} \breve{\va}_{i}^\tr$.

\item {\em Form calibration grid:}  Repeat steps \ref{calStep2}--\ref{calStep4} in a grid pattern corresponding to the desired distribution of beam centers and constraint points, e.g.\  for spherical angles $\{\Omega_i \, | \, 1 \leq i \leq K \}$.

%

\end{enumerate}

\subsection{Beamformer Design}
\label{sec:Beamformers}

In order to form beams with a PAF, a design procedure is required to obtain a set of beamformer weight vectors $\vw_{i}^{(j)}$, from which the output for the $i$th beam is computed as
\begin{equation}
y_i [n] = {\vw_{i}^{(j)\,H}} \vx [n],  \;\;\; 0 \leq i \leq K,  \;\;\; j = \left\lfloor {\frac{n}{L}} \right\rfloor
\label{eqn:beamformer}
\end{equation}
A distinct beamformer weight vector $\vw_{i}^{(j)}$ is used for each main lobe steering angle $\Omega_i$.  For beamformer weights that are fixed during each STI, the dependence on the STI index $j$ can be dropped.  In a practical PAF observing scenario the beams are steered in a rectangular or hexagonal grid pattern with crossover points at the -1 to -3 dB levels.  Proper design of a set of beamformer weights $\vw_i$ allows one to steer the main beams, control beam shape and sidelobe levels, optimize sensitivity, and direct placement of nulls towards interferers.  A thorough introduction to array signal processing and beamformer design can be found in \citep{vanveen1988,VanTrees02IV}.

In general, beamformer weights can be designed using an a priori electrical model for the array, characterization data measured before installation of the array, or in situ calibration data obtained with a procedure such as that outlined in the previous section.  As will be argued below, only the latter of these beamformer design procedures is likely to be viable for astronomical PAFs.

Relative to phased arrays for wireless, communications, radar, sonar, and other applications, an important distinguishing characteristic of astronomical PAF beamforming is that given the current state of the art the instrument performance with data independent pre-computed beamformer weights is inadequate.  For simulation studies, the conjugate field match beamformer has been commonly used to obtain beamformer weights, but the only demonstrated experimental radio camera images formed to date use statistically optimal, adaptive, data dependent beamforming algorithms \citep{apertif_08,Oosterloo08,Jeffs08c,paf_warnick_ursi08,Jeffs09}.
The primary reason for this is that the perturbations in array element patterns, complex receiver gains, and cable phase lengths that occur when the array is mounted on the reflector mean that numerical models or antenna range characterization are not accurate enough to allow precomputation of beamformer weights for high sensitivity or controlled pattern shapes.

In applications with less stringent sensitivity and stability requirements, it is possible to
design 2-D array beamformers deterministically to match some fixed array response criterion.
A number of classical methods are used, including windowed beamforming and numerical response optimization over a dense grid of far field sample points \citep{vanveen1988,VanTrees02IV}.
It has been shown in simulation that using dense calibration response and numerical optimization it is possible to design well behaved astronomical PAF beam shapes with little coma, distortion, or polarization rotation as beams are steered off-axis \citep{Willis09}.
All such methods require an accurately calibrated array (by antenna range measurements or mathematical modeling) over the entire response field of interest.

With astronomical PAF instruments it is not possible a priori to obtain the full-sphere calibration data required to design beamformer weights for a given response criterion.  Antenna range measurements of the bare PAF element response patterns do not account for such effects as interactions with support structures.
Numerical modeling of the array and dish combination provides an excellent qualitative match, but our studies of precomputed beamformer weights indicate that modeled results differ sufficiently in fine phase and gain detail over the field of view.
Thus beamformer weights designed using numerical electromagnetic simulations lead to on-dish real data beamformer results with lowered sensitivity and beampatterns that deviate significantly from the desired shape.

Detailed modeling of the array elements, backplane, array structure, and dish reflector system can provide excellent qualitative match and prediction of achievable real-world performance when used to compute beampatterns in simulation.
The fine detail mismatch though precludes use of these beamformer weight designs directly on the real system.
We speculate that the discrepancies are in subtle fine scale differences between physical geometry and the model.
The simulations can predict how {\em an} array and dish of this design will perform, but cannot match the exact phase and gain variation subtleties of a given telescope necessary to design a controlled beampattern.
Details of reflector surface roughness, support structure scattering, small variations in element position and construction, noise temperature variation between LNAs, etc.
can only be captured statistically in the models, but the exact unknown values cannot be matched to the real values for a specific dish and array.

Since a priori calibration is insufficiently accurate, in situ calibration is required.
Even with bright calibrator sources, due to limited SNR and integration time requirements it is not practical to obtain calibration vectors $\va_i$ with adequate density or coverage beyond the first one or two sidelobes (although correlation of the array outputs with a second high sensitivity antenna could be used to improve coverage).  Errors introduced in transferring a beamformer design from simulation invariably lead to unacceptable reduction in real-world sensitivity.  Further, fine control of beam pattern shape to reduce temporal variation of the beam response (``pattern rumble'') and insure uniformity across all pixels cannot be attained with pre-computed data independent designs.  We have investigated these deterministic design methods, but performance using real experimental data is poor.  We are currently studying methods for translating a desirable beamformer design from a detailed numerical simulation to the corresponding actual PAF array using available in situ calibrations, but we have found that simply correcting for the per-channel phase and gain calibration is insufficient due to the effects of array mutual coupling.

For these reasons, rather than designing the beam pattern shape over a full sphere using classical beam pattern fitting methods, we have proposed using statistically optimal, data-dependent beamforming design methods for PAF array beamforming \citep{Jeffs08b}.  This approach uses calibration grid data over a limited field of view near the reflector boresight together with the measured noise response of the array to design beamformer weight coefficients.  In effect, the calibration grid data controls the pattern main lobe for each beam, and the measured array noise response is used to optimize the deep sidelobes to minimize spillover noise.  Since receiver noise for a PAF is correlated, the statistically optimal beamformer design approach also suppresses receiver noise to a degree as well.

We have used two of the several well known statistically optimal beamforming algorithms: the linearly constrained minimum variance (LCMV) and maximum sensitivity (max SNR) beamformers \citep{vanveen1988,VanTrees02IV}.  LCMV minimizes total output power while satisfying a set of linear constraints. The LCMV optimization problem is given by
	\begin{equation}
	\label{eq:LCMV_prob}
	\vw_{{\rm LCMV},i}  =  \arg \min_\vw \vw^H {\mR} \vw \hbox{ \ \ subject to \ \ } \vw^H \mC_i = \vf
	\end{equation}
For example, using  $\mC_i  =  [ \va_{i}, \va_{j_1} , \cdots, \va_{j_P} ]$ and $\vf = [ 1, r_1, \cdots, r_P]^T$ constrains the beam main lobe to have a response of unity in direction $\Omega_i$ and $r_p$ in $P$ other directions, $\Omega_{j_p}$.
The single constraint minimum variance distortionless response (MVDR) beamformer is a special case of LCMV obtained when $\mC_i = \va_i$, and $\vf = 1$.

Using calibration data only, (\ref{eq:LCMV_prob}) can be solved approximately as
	\begin{equation}
	\vw_{{\rm LCMV},i} = \breve{\mR}_{\rm n}^{-1} \mC_i [ \mC_i^H \breve{\mR}_{\rm n}^{-1}\mC_i]^{-1} \vf
	\label{eq:LCMV_wts}
	\end{equation}
We refer to this solution as a ``fixed--adaptive'' beamformer since though it is statistically optimal for the calibration noise environment, it is computed using a noise covariance measured before or after the observation phase and and does not use the current array covariance during the observation. For noise field tracking or moving interference canceling we use a fully adaptive mode, replacing $\breve{\mR}_{\rm n}$ in Eq.\ (\ref{eq:LCMV_wts}) with $\hat{\mR}^{(j)}$ and updating the weight computation for each STI.

The maximum sensitivity beamformer is defined as
\begin{equation}
\label{eq:maxSNR_prob}
\vw_{{\rm mSNR},i} = \arg \max_\vw \frac{\vw^H {\mR}_{\rm s} \vw}{\vw^H {\mR}_{\rm n} \vw } .
\end{equation}
The maximization in Eq.\ (\ref{eq:maxSNR_prob}) invokes the generalized eigenvector problem
\begin{equation}
\breve{\mR}_{{\rm s},i} \vw_{{\rm mSNR},i} = \lambda_{\rm max} \breve{\mR}_{\rm n} \vw_{{\rm mSNR},i}
\end{equation}
for which the solution using only calibration data yields a practical fixed-adaptive beamformer.
Rapid tracking to adapt to noise field evolution can be accomplished by replacing calibration $\breve{\mR}_{\rm n}$ in Eq.\ (\ref{eq:maxSNR_prob}) with periodically updated ``off-source-steered'' estimates $\hat{\mR}_{\rm off}$ obtained during observation.
This method is not well suited to interference mitigation since off-steering changes the spatial structure of $\mR_{\rm int}$ in $\mR_{\rm n}$.  The maximum sensitivity beamformer has been used for most experimental PAF observations reported to date \citep{paf_warnick_ursi08,apertif_08,Oosterloo08, Jeffs08c, Jeffs09}.

Some of the PAF science applications described in Section \ref{sec:intro} require beams with very low sidelobes.  The maximum sensitivity beamformer \eqref{eq:maxSNR_prob} does not result in lowest possible sidelobes.  In principle, the same calibration grids used to generate the signal response correlation matrices $\breve{\mR}_{{\rm s},i}$ can also be used to design beams with controlled sidelobes.  By definition, these beams will result in lower sensitivity than \eqref{eq:maxSNR_prob}, but the overall instrument performance should be better for observations requiring low sidelobes.  We are currently investigating beamformer design procedures that incorporate beam shape metrics while still maintaining as high a sensitivity as possible.  The flexibility to optimize beams for different science applications in post-processing represents an advantage of PAFs over fixed feeds.
Sidelobe pattern ``rumble'' or variation between dishes is of particular concern in PAF synthesis imaging as is planned for the ASKAP array and PAF upgrade to the WSRT.
Stable know sidelobe patterns permit high dynamic range imaging.
Since calibration grids are only possible (due to SNR limitations) out to the first sidelobe or two, it will not be possible to strictly control deep sidelobe patters with the beamformer weights.
On the other hand, outside the first few sidelobe rings, the highly attenuating pattern of the dish aperture itself dominates over anything the array feed can control.
This is a positive promising aspect of sidelobe pattern rumble control that needs further study.


\section{Performance Metrics and Noise Models}
\label{sec:noiseModels}

\newcommand{\wv}{{\bf w}}
\newcommand{\Rm}{{\bf R}}

\subsection{Sensitivity and Efficiency}

For a phased array, sensitivity, efficiencies, and system noise temperature are beam-dependent.  The primary figure of merit for a formed beam in the absence of interference is the beam sensitivity
\begin{equation}
   \label{eq:sensitivSNR}
    \frac{A_{\rm eff}}{T_{\rm sys}} =
    \frac{2 k_{\rm b}}{10^{-26} F^{\rm s}} {\rm SNR} \; = \;
    \frac{2 k_{\rm b}}{10^{-26} F^{\rm s}} \frac{ \vw^\tr \mR_{\rm s} \vw}{\vw^\tr \mR_{\rm n} \vw}
\end{equation}
where $A_{\rm eff}$ is the effective receiving area for the PAF illuminated dish using beamformer weight $\vw$, $T_{\rm sys}$ is the beam equivalent noise temperature, $k_{\rm b}$ is Boltzmann's constant, and $F^{\rm s}$ (Jy) is the flux density of a signal source of interest.
Sensitivity is related to radiation, aperture, and spillover efficiencies ($\eta_{\rm rad}$, $\eta_{\rm ap}$, and $ \eta_{\rm sp}$ respectively) according to
\begin{equation} \label{eq:sensitiv}
    \frac{A_{\rm eff}}{T_{\rm sys}}
    =
    \frac{\eta_{\rm rad} \eta_{\rm ap} A_{\rm p}}
    {\eta_{\rm rad}(T_{\rm sp} + T_{\rm sky}) 
     + (1-\eta_{\rm rad}) T_{\rm p} + T_{\rm rec}}
\end{equation}
where $A_{p}$ is the physical aperture area; $T_{\rm sp}$ is the equivalent spillover noise temperature; $T_{\rm sky} = T_{\rm cmb} + T_{\rm gb} + T_{\rm atm} $ is the combined non-SOI noise temperature in the beam main lobe due to cosmic microwave background radiation, galactic background, and atmospheric noise; $T_{\rm rec}$ is the beam equivalent receiver noise temperature; and $T_{\rm p}$ is the physical temperature of the array antenna.  These constituent noise temperatures arise from the corresponding array noise covariance terms $\mR_{\rm sp}$, $\mR_{\rm sky}$, $\mR_{\rm loss}$, and $\mR_{\rm rec}$.


Defining antenna figures of merit is straightforward for a passive antenna, but for an active receiving array the standard definitions for antenna terms are not directly applicable.  The IEEE standard definition of aperture efficiency can be extended to active arrays using \citep{effic_warnick_08}
\begin{align}\label{eq:eta_ap}
    \eta_{\rm ap} = \frac{2 k_{\rm b} T_{\rm iso} B}{ 10^{-26} F^{\rm s} A_{\rm p}}
    \frac{\vw^{\tr} \mR_{\rm s} \vw}{\vw^{\tr} \mR_{\rm iso} \vw}
\end{align}
where $\mR_{\rm iso}$ is the correlation matrix of the array output voltages due to an isotropic external noise field at temperature $T_{\rm iso}$ (i.e., the correlation matrix of the array noise output with lossless antenna and noiseless receivers and with the antenna in an isotropic thermal environment), $B$ is the system noise equivalent bandwidth, and $\mR_{\rm s}$ is the array output voltage correlation matrix due to a signal of interest.
A method for measuring   $\mR_{\rm iso}$ experimentally is described in Section \ref{sec:hotCold}.

\subsection{Noise Models}
\label{sec:noise_model}

\begin{figure}[tb]
\centering
\includegraphics[width = 3in]{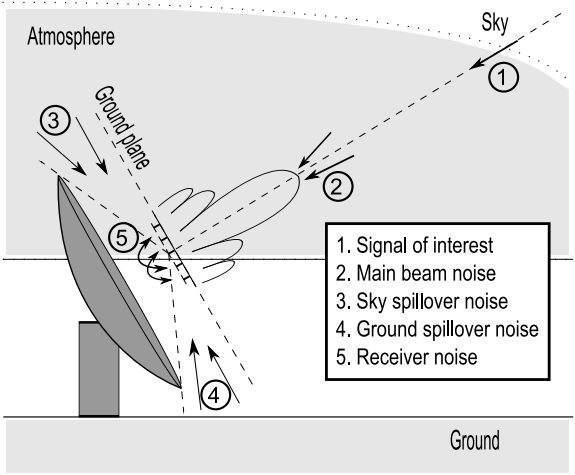}
\caption{The sources of noise that contribute to the signal received by a radio telescope include spillover noise originating from outside the edge of the reflector dish, atmospheric noise collected by the main lobe of the antenna radiation pattern, and receiver noise.  The equivalent receiver noise temperature includes the effects of mutual coupling between antenna array elements and impedance mismatches between the array and receivers.}
\label{fig:full_noise}
\end{figure}


The beam equivalent system noise temperature is \citep{effic_warnick_08}
\begin{align}
    T_{\rm sys} & = \eta_{\rm rad} (T_{\rm sky} + T_{\rm sp}) + T_{\rm loss} + T_{\rm rec}
    \nonumber \\
    & = \eta_{\rm rad} T_{\rm iso} \frac{ \vw^\tr ( \mR_{\rm rec} + \mR_{\rm sp} + \mR_{\rm sky} + \mR_{\rm loss} ) \vw}{\vw^\tr \mR_{\rm iso} \vw} \nonumber \\
    & = T_{\rm iso} \frac{ \vw^\tr \mR_{\rm n} \vw}{\vw^\tr (\mR_{\rm iso} + \mR_{\rm loss}) \vw}
    \label{eq:Tsys}
\end{align}
The following sections present analytical and numerical models used to simulate the noise covariance matrices introduced in Eq.\ (\ref{eq:noiseCovars}), used in Eq.\ (\ref{eq:Tsys}) and illustrated in Fig.\ \ref{fig:full_noise}.

\subsubsection{Receiver Noise}
Because antenna elements of a PAF are closely spaced, mutual coupling between elements is strong.  This influences the equivalent receiver noise temperature through coupling of noise emitted by the input ports of low-noise amplifiers (LNAs), as described by \cite{warnick2005}.  To account for these effects, the receiver noise covariance matrix $\mR_{\rm rec}$ is computed in the sequel according to the model \citep{mc_warnick_08}
	\begin{equation}
	\mR_{{\rm rec}} = 2B\mQ[\mV^2_{\rm n} + \mZ_{\rm ar}\mY_{\rm c}\mV^2_{\rm n} + \mV^2_{\rm n}\mY^H_{\rm c}\mZ^H_{\rm ar} + \mZ_{\rm ar}\mI^2_{\rm n}\mZ^H_{\rm ar}]\mQ^H
	\label{eq:Rrec}
	\end{equation}
where $\mZ_{\rm ar}$ is the array impedance matrix, and $\mQ$ is defined as
	\begin{equation}
	\mQ = \mZ_{\rm rec}(\mZ_{\rm rec}+\mZ_{\rm ar})^{-1}
	\end{equation}
where $\mZ_{\rm rec}$ is the load impedance matrix.  $\mV_{\rm n}$, $\mI_{\rm n}$, and $\mY_{\rm c}$ are diagonal matrices of LNA noise parameters (noise voltage densities, noise current densities, and correlation admittances, respectively).  As described by \cite{mc_warnick_08}, these terms depend on the minimum achievable (at perfect impedance match) equivalent temperature $T_{\rm min}$ associated with each LNA.
In the absence of mutual coupling, $\mR_{\rm rec}$ is diagonal, and each LNA can be noise matched to the array ports such that for all beams
\begin{equation}
T_{\rm rec} =  T_{\rm iso} \frac{ \vw^\tr \mR_{\rm rec} \vw}
      {\vw^\tr (\mR_{\rm iso} + \mR_{\rm loss}) \vw} \; = \; T_{\min}
\end{equation}
where we have neglected downstream noise in the receiver chains.  When LNA noise couples back through array elements to neighboring closely packed antennas, off diagonal terms in $\mR_{\rm rec}$ are non--zero, and $T_{\rm rec}$ is increased by a mutual coupling noise penalty as seen in Section \ref{sec:hotCold}.  Because the equivalent receiver noise temperature is referred to the reference plan at the array element ports, it includes the effects of mismatch loss at the interface between the array and LNAs.

\subsubsection{Spillover Noise}
\label{sec:spillover}


This section presents a numerical model for spillover noise.  The PAF beamformed illumination response pattern extends beyond the edge of the reflective surface, collecting undesired signals arriving from the spillover region.  With a ground plane backing the array, backlobes are relatively small, so the spillover region is assumed to extend from the dish edge to the plane of the array.

For a given reflector tipping angle, the spillover region includes both sky noise and relatively high temperature (e.g., $280\,$K) ground.  To account for this temperature distribution, we have modeled spillover noise as a dense grid of independent point sources with approximately uniform angular spacing on a spherical annulus from the perspective of the PAF.  The center is obscured by the primary reflector.  As the dish tips, a portion of the annulus rises above the extended horizon plane, corresponding to the part of the spillover illumination pattern observing cold sky rather than warm ground.



The spillover noise covariance matrix is approximately
    \begin{equation}
    \mR_{\rm sp} = \frac{16 k_{\rm b} B}{\mid{I_0}\mid^2}\frac{1}{2\eta}
    \sum_i{T_i \mQ \va_i\va_i^H \mQ^H \alpha_i}
	\label{eq:Rspill}
    \end{equation}
where $I_0$ is the element excitation input current, $\eta$ is the intrinsic impedance of space, $T_i$ is the noise temperature associated with the $i$th grid point, $\va_i$ is the array response vector due to a unit amplitude source at the $i$th spillover noise grid position, and $\alpha_i$ is the solid angle of the corresponding sky patch.  
For grid points below the horizon, $T_i = 280\,$K, representing warm ground.
Above the horizon correspond, the sky noise model $T_i = T_{\rm atm}(\theta_i) + T_{\rm cmb} + T_{\rm gb}$ is used, where $T_{\rm atm}(\theta)$ is the elevation dependent atmospheric noise model developed below in
Eq.\ (\ref{eqn:T_atm}) and $\theta_i$ is the zenith angle.

\subsubsection{Sky Noise}
\label{sec:main_beam}

Atmospheric noise, CMB, and galactic background (GB) seen though the beamformer main lobe in the observation pointing direction all contribute to sky noise. Outside the main beam, these noise sources are attenuated by the telescope's sidelobe pattern and can be neglected in the total system noise model.  Atmospheric noise increases as the dish is tipped toward the horizon, while CMB and GB noise are modeled as a constant at all elevations.  As will be shown in Section~\ref{sec:simresults}, sky noise becomes the dominant source as the antenna boresight approaches the horizon.

An extended thermal noise source larger than the main beam can be extended to an isotropic (i.e., full sphere) noise distribution with only a small perturbation to the antenna response.  We can therefore use the approximate model to represent sky noise:
    \begin{eqnarray}
	\mR_{\rm sky} & = & \frac{T_{\rm sky}}{T_{\rm iso}} \mR_{\rm iso} 
	\label{eq:Rmb} \\
	T_{\rm sky} & = & T_{\rm atm} + T_{\rm cmb} + T_{\rm gb}  \nonumber
	\end{eqnarray}
The isotropic noise correlation matrix $\mR_{\rm iso}$ can be computed from the array element pattern overlap integral matrix defined by \cite{warnick2006}.  Though the SOI is also seen in the main lobe, its contribution is contained in $\mR_{\rm s}$ while $\mR_{\rm sky}$ includes only noise terms.

\label{sec:sky_temp}


For atmospheric noiase, we use a modified plane-parallel atmosphere model.  At a given elevation angle, $T_{\rm atm}$ is proportional to the line--of--sight thickness of the atmosphere, which for simplicity is assumed to be a solid slab of uniform thickness on a flat earth surface.
The path length through the atmosphere increases with depression angle $\theta$ according to $d(\theta) = d_0 \sec(\theta)$, where $d_0$ is the distance corresponding to the zenith direction.
Since $T_{\rm atm}$ seen in the beam main lobe is approximately proportional to the corresponding $d(\theta)$, we have
	\begin{equation}
	 \label{eqn:T_atm}
	T_{\rm atm}(\theta) = \left\{ \begin{array}{l r}
	T_{\rm 0,atm}\sec(\theta) & 0 \leq \theta \leq 80^\circ \\
	T_{\rm 0,atm}\sec(80^\circ)\\ \;\;\; + 1.3\,(\theta-80^\circ) & 80^\circ < \theta \leq 90^\circ
	\end{array} \right.
	\end{equation}
where $T_{\rm 0,atm}= 2$K is the temperature at zenith.
Note that for depression angles greater than $80^\circ$ a correction is included to avoid the infinity at the horizon \citep{roddy2006}.
We assume an isotropic brightness temperature distribution for cosmic background and galactic background noise, with $T_{\rm cmb} + T_{\rm gb} = 3$K.

\subsubsection{Loss Noise}
\label{sec:loss}
Resistive losses in the antenna, cables, and connectors ahead of the LNA introduce a noise source that is difficult to model or measure accurately.  We adopt the approximate diagonal covariance matrix model
\begin{equation}
\mR_{\rm loss} = \sigma_{\rm loss}^2 \, \mI
\end{equation}
For a given value of $T_{\rm loss}$,
\begin{equation}
\sigma_{\rm loss}^2 = \frac{T_{\rm loss}}{T_{\rm iso}}  \frac{\vw^{\tr} \vw}{\vw^{\tr} \mR_{\rm iso} \vw}
\end{equation}
In the 19 element PAF, the dominant source of loss noise is a short length of coaxial line feeding each array element.  From measurements of the cable loss, we have obtained the estimate $T_{\rm loss} \simeq 5$K.


\section{Experimental Results}

\subsection{Noise and Sensitivity Measurements}
\label{sec:hotCold}

The isotropic noise correlation matrix used to compute beam aperture efficiencies and system noise temperatures was measured using the ground shield facility shown in Fig.~\ref{fig:SkyNoiseFacility} and described in Section \ref{sec:ground_shield}.  The array output correlation matrix $\breve{\mR}_{\rm cold}$ for an isotropic cold source (sky) and $\breve{\mR}_{\rm hot}$ with the array aperture covered by microwave absorber at ambient temperature were acquired for the PAF system at 1600 MHz.  Using an array generalization of the $Y$-factor technique, the isotropic noise response of the array can be obtained from Eq.~\eqref{eq:Tsys} in the form
\begin{equation}\label{eq:Riso}
   \breve{\mR}_{\rm iso} = \frac{T_{\rm iso}}{T_{\rm hot} + T_{\rm cold}} (\breve{\mR}_{\rm hot} - \breve{\mR}_{\rm cold})
\end{equation}

With the PAF mounted on the reflector, the array response correlation matrices for on-source, $\breve{\mR}_{{\rm s},i}$, and off-source pointings, $\breve{\mR}_{\rm n}$, were obtained for each beam steering direction $\Omega_i$ using calibration data as described in Section  \ref{sec:Cal}.  These are combined with $\breve{\mR}_{\rm iso}$ and the known calibrator flux density $F^{\rm s}$ to compute estimates of beam sensitivities, aperture efficiencies and system temperature using (\ref{eq:sensitivSNR}), \eqref{eq:eta_ap}, and (\ref{eq:Tsys}) respectively.

To validate the experimental measurements, a finite element method (FEM) numerical model of the array was created using HFSS (Ansoft Corp.).  The ground plane and the dipole elements are modeled as perfect electric conductors, and a loss term is added to account for the measured loss of the coaxial cable from the dipoles to the element terminal connectors.

\begin{table}[!t]
\renewcommand{\arraystretch}{1.3}
\caption{Measured and modeled peak beam sensitivity, system temperature, and aperture efficiency for the 19 element prototype dipole array.}
\label{tab:sens}
\centering
\begin{tabular}{llll}
& Center Element & Formed Beam & Model \\ \hline


Sensitivity & 2$\,$m$^2$/K & 3.3$\,$m$^2$/K & 3.7$\,$m$^2$/K \\
$T_{\rm sys}$ & 101 K & 66 K & 69 K \\

$\eta_{\rm ap}$  & 64\% & 69\% & 81\%
\end{tabular}
\end{table}

\begin{table}[!t]
\renewcommand{\arraystretch}{1.2}
\caption{System noise budget.  }
\label{tab:noise}
\centering
\vspace{.1in}
\begin{tabular}{lccc}
& \centering{Measured} & Modeled  \\ \hline
LNA $T_{\rm min}$ & 33 K & 33 K \\
Mutual coupling   & 20 K & 23 K \\
Spillover         &  5 K &  5 K \\
$T_{\rm sky}$               &  3 K &  3 K \\
$T_{\rm loss}$              &  5 K &  5 K \\ \hline
$T_{\rm sys}$     & 66 K & 69 K
\end{tabular}
\end{table}

Table \ref{tab:sens} shows measured and modeled results for the 19 element prototype array for the sensitivity, aperture efficiency, and system temperature for the center element beam (i.e., $\vw^T = [1 \,\, 0 \,\, \ldots \,\, 0]$) and the full-array formed beam with peak sensitivity.  The 12\% discrepancy between measured and modeled aperture efficiency is expected, given that blockage and feed support scattering are not modeled and that numerical radiation patterns for any type of feed typically overestimate efficiency by roughly 10\% \citep{thesis_murphy}.


The system noise temperature for a formed beam was estimated using Eq.~\eqref{eq:Tsys} expressed in the form
\begin{equation}
  T_{\rm sys} \simeq \eta_{\rm rad} T_{\rm iso}
  \frac{\vw^\tr \hat{\mR}_{\rm off} \vw}
  {\wv^\tr \breve{\mR}_{\rm iso} \vw}
\end{equation}
The isotropic noise response essentially provides the available receiver gain $\wv^\tr {\mR}_{\rm iso} \vw/(\eta_{\rm rad} k_b T_{\rm iso} B)$ for the formed beam, allowing the output noise power to be expressed as an equivalent sky temperature.  This array characterization technique is particulary convenient for phased arrays, but is influenced by nonuniformities in the hot and cold noise sources and receiver gain variations when the PAF system is moved from the warm absorber/cold sky facility to the reflector.  With a horizon model to determine the sky brightness temperature nonuniformity and comparing measured results to numerical models, the temperatures obtained using this technique appeared to be accurate to within roughly 10\%.

The system noise budget for a formed beam is shown in Table \ref{tab:noise} for the 19 element array.  Noise contributions due to mutual coupling and spillover were not separately measured, but were estimated by carrying over the modeled spillover noise and adjusting the mutual coupling contribution to make up the total measured $T_{\rm sys}$.  The dominant source of loss was the coaxial feed for each dipole.

The mutual coupling term in Table \ref{tab:noise} is a contribution to the equivalent receiver noise temperature $T_{\rm rec}$ caused by impedance mismatches between the LNAs and the array.  The total receiver noise temperature for the formed beam is $T_{\rm rec} = 53\,$K.  The optimum source impedances of the front end amplifiers were matched to the isolated input impedances of the array elements, but due to mutual coupling between array elements the effective impedance presented to each front end amplifier is an active impedance that is different from the element self impedance \citep{active_woestenburg_tr_05}.  The resulting mismatches between the optimum source impedance parameter of the LNAs and the active impedances presented by the array to the amplifiers led to an increase in the equivalent receiver noise.  The active impedances are beamformer-dependent, which means that the equivalent receiver noise for a phased array varies with respect to the beam scanning angle.

Improving the noise match between the front end amplifiers and the array active impedances to reduce the mutual coupling contribution to the system noise is a major focus of current work in PAF development.  For communications applications, decoupling networks have been explored \citep{opt_match} but are likely too lossy and narrowband for astronomical instruments.  More promising approaches include a noise matching condition that is optimal in an average sense over the array field of view \citep{PAF_iwat09_warnick,mc_warnick_08}, and the design of array antennas with active impedances that remain as close as possible to the LNA optimum source impedance (e.g., 50$\,\Omega$) over the array field of view and operating bandwidth.  Results on the expected performance improvement with active impedance matching are given in Sec.\ \ref{sec:tipping_experimental}.

\begin{figure}[!t]
\begin{center}
\subfigure[]{
\includegraphics[width=2.9in]{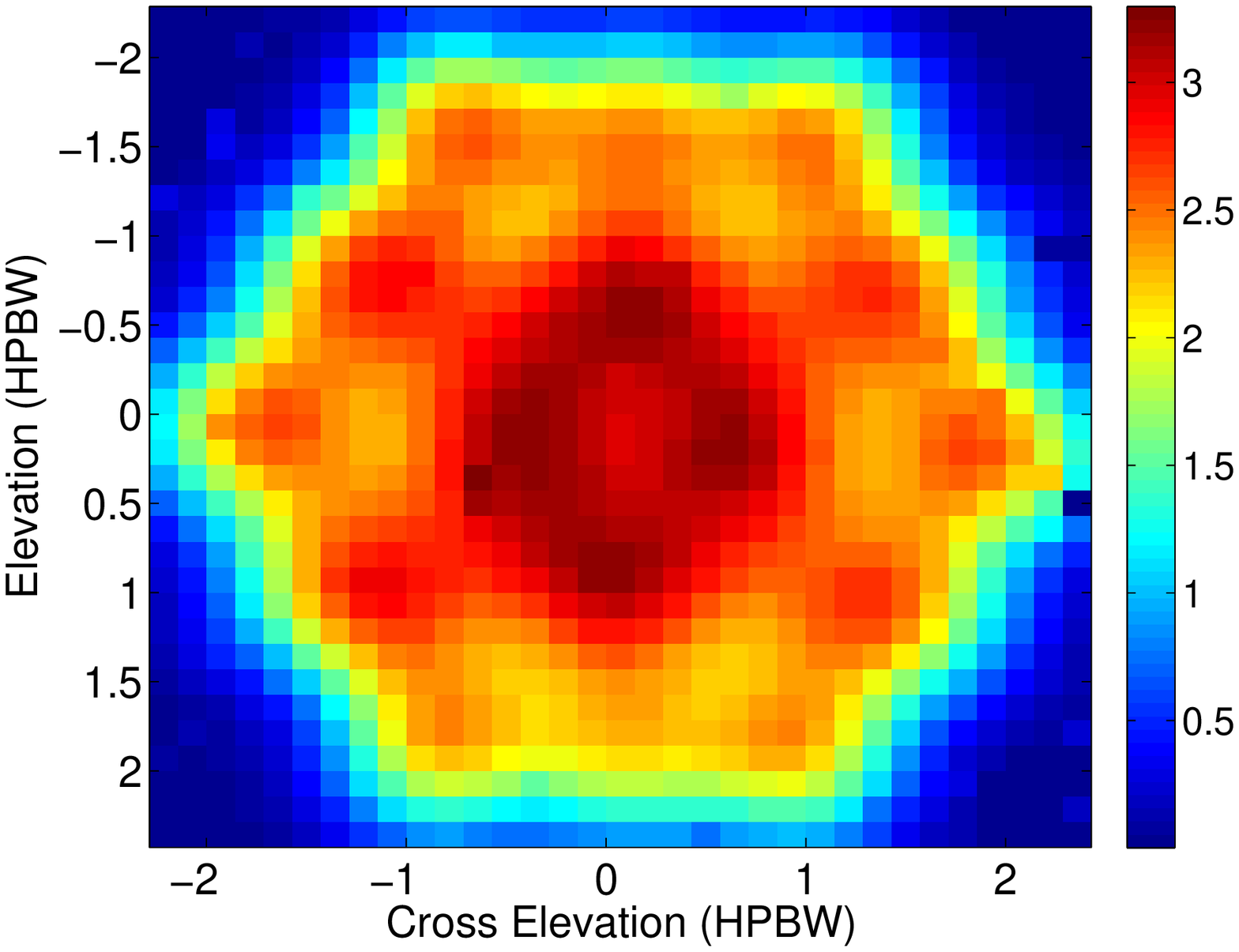}
}
\subfigure[]{
\includegraphics[width=2.5in]{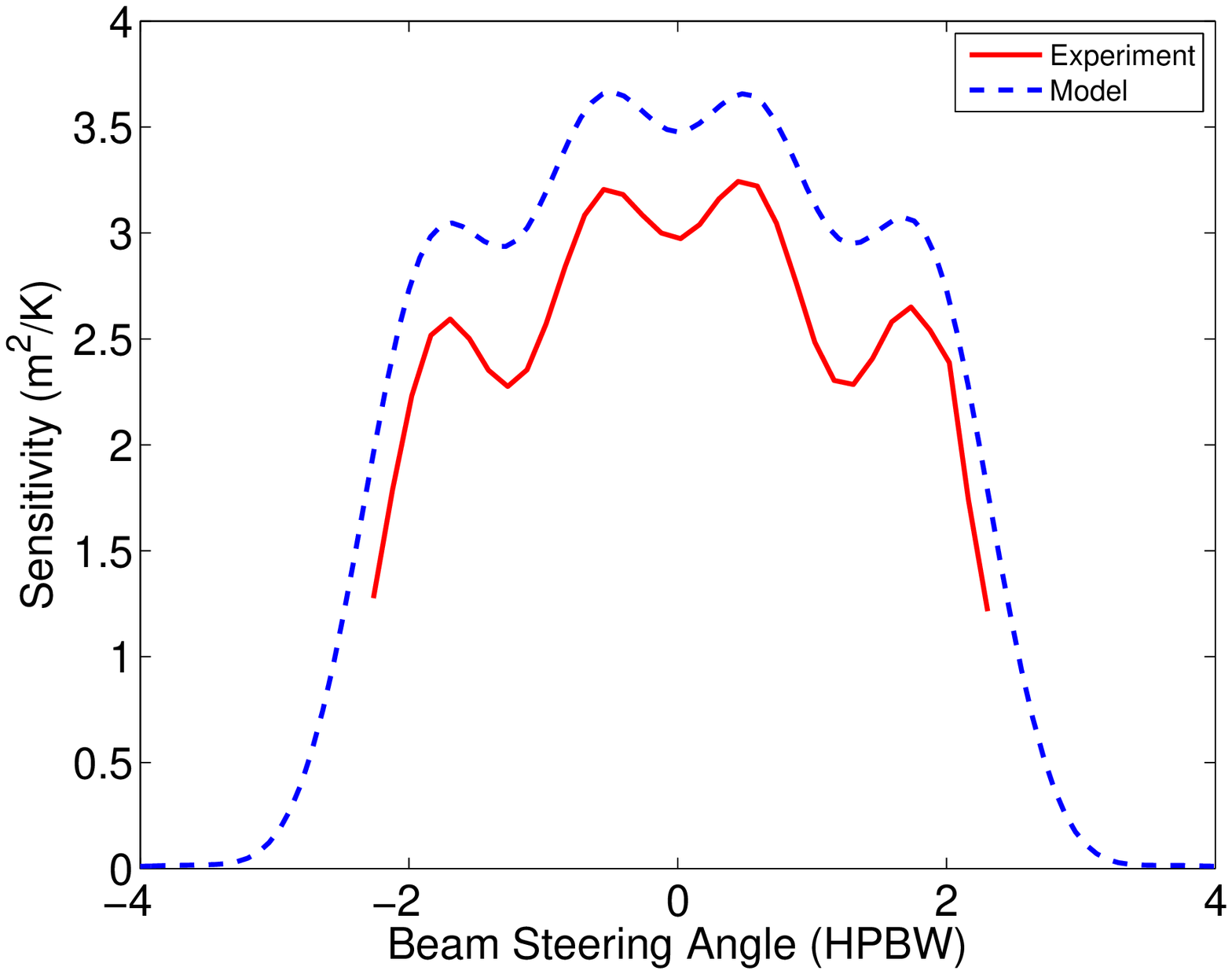}
}
\caption{(a) Measured beam sensitivity map (m$^2$/K) for the 19 element dipole array on the 20-Meter Telescope ($f/D = 0.43$).  Each pixel in the image corresponds to the measured sensitivity of one formed beam.  (b) Measured and modeled beam sensitivity for an elevation cut through the PAF field of view.  Measured sensitivities were obtained using a bright calibrator source on a grid of reflector steering directions with 0.1 degree spacing. The model employed HFSS for the array, physical optics for reflector scattering, and a microwave network model for the receiver chains and beamformer. The half-power beamwidth (HPBW) is 0.7 degrees.  }
\label{fig:sens}
\end{center}
\end{figure}

The values given above are for a single beam steered near the reflector boresight direction.  All figures of merit, including aperture efficiency and system temperature as well as sensitivity, vary across the PAF field of view.  The antenna figures of merit can be mapped using data acquired for a grid of calibrator source pointings.  For each beam, an off-source pointing and an on-source pointing with the reflector steered so that the beam is centered on the source were used to compute SNR, which can be converted to sensitivity using the known source flux density.  A map of beam sensitivities is shown in Fig.~\ref{fig:sens}(a).
In this and all images in Section \ref{sec:RadioCamera} the horizontal axis is given in ``cross elevation'' rather than azimuth, in order to reduce coordinate system projection distortion on the sky grid.  At the image midpoint, cross elevation describes an arc perpendicular to the arc running from horizon to zenith.
This provides an undistorted, uniform pixel size rectilinear grid projection on the sky.  Near the horizon, cross elevation is equivalent to azimuth.  For an image centered at zenith, both elevation and cross elevation axes follow arcs from zenith to horizon.

Figure \ref{fig:sens}(b) shows a slice through the sensitivity map with a comparison to the model results.  The shape of these sensitivity maps depends on many factors, including the size of the PAF, the number of elements used to form a beam (in this case, all 19 elements were used for each beam), and the location of the focal spot in relation to the array elements.
Based on a 1 dB tolerance for sensitivity loss for steered beams and a raster spacing with beam overlap at the half-power point, the measured field of view is 2.5 HPBW in diameter.  These results show good performance for the prototype PAF and demonstrate that scientifically useful sensitivities and efficiencies can be obtained.

\subsection{Radio Camera Imaging}
\label{sec:RadioCamera}

The primary motivation for PAF telescope instruments is achieving wide fields of view with multiple, electronically steered beams forming a radio image with a single dish pointing.  As part of the July 2008 experimental campaign on the Green Bank 20-Meter Telescope we collected a number of calibration data grids, used these to compute simultaneous beamformer weights using the maximum sensitivity beamformer method of (\ref{eq:maxSNR_prob}), and observed several astronomical radio sources with the PAF operating as a radio camera.  Calibration grids were either 33 $\times$ 33 or 65 $\times$ 65 pixels in size, with inter-pixel separation of $0.1^\circ$, where each pixel represents a calibration pointing direction as defined in Section \ref{sec:Cal}.  For diagnostic purposes, the calibration grids were larger than the PAF field of view, so a subset of the grids was used to form images.  One beam was formed per calibration pixel.  When observing a region larger than the PAF field of view, the reflector was physically steered over a grid of pointings with approximately $1.0^\circ$ spacing in order to form a mosaic of PAF images.

\begin{figure}[!t]
\centering
\includegraphics[width=3.0in]{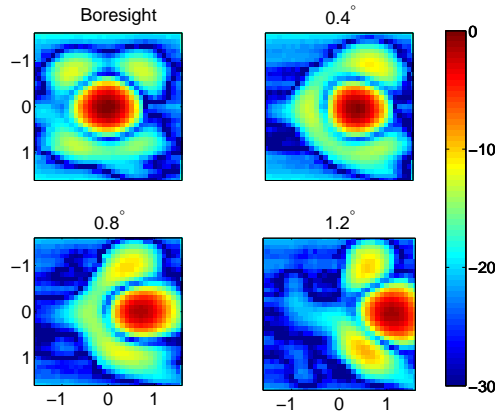}
\caption{Measured beam receiving patterns (dB relative to peak).  The 1.2$^\circ$ beam is beyond the PAF field of view as defined by a 1$\,$dB sensitivity loss.}
\label{fig:beams}
\end{figure}

Receiving patterns for several of the resulting beams are shown in Fig.~\ref{fig:beams}.  The peak sidelobe level was better than 10 dB for beams steered up to $0.6^\circ$, which is near the edge of the PAF field of view as defined by a 1$\,$dB sensitivity loss.


\begin{figure}[!t]
\centering
\subfigure[Single reflector pointing, multibeam image of 3C295.]{
\includegraphics[width=3.0in]{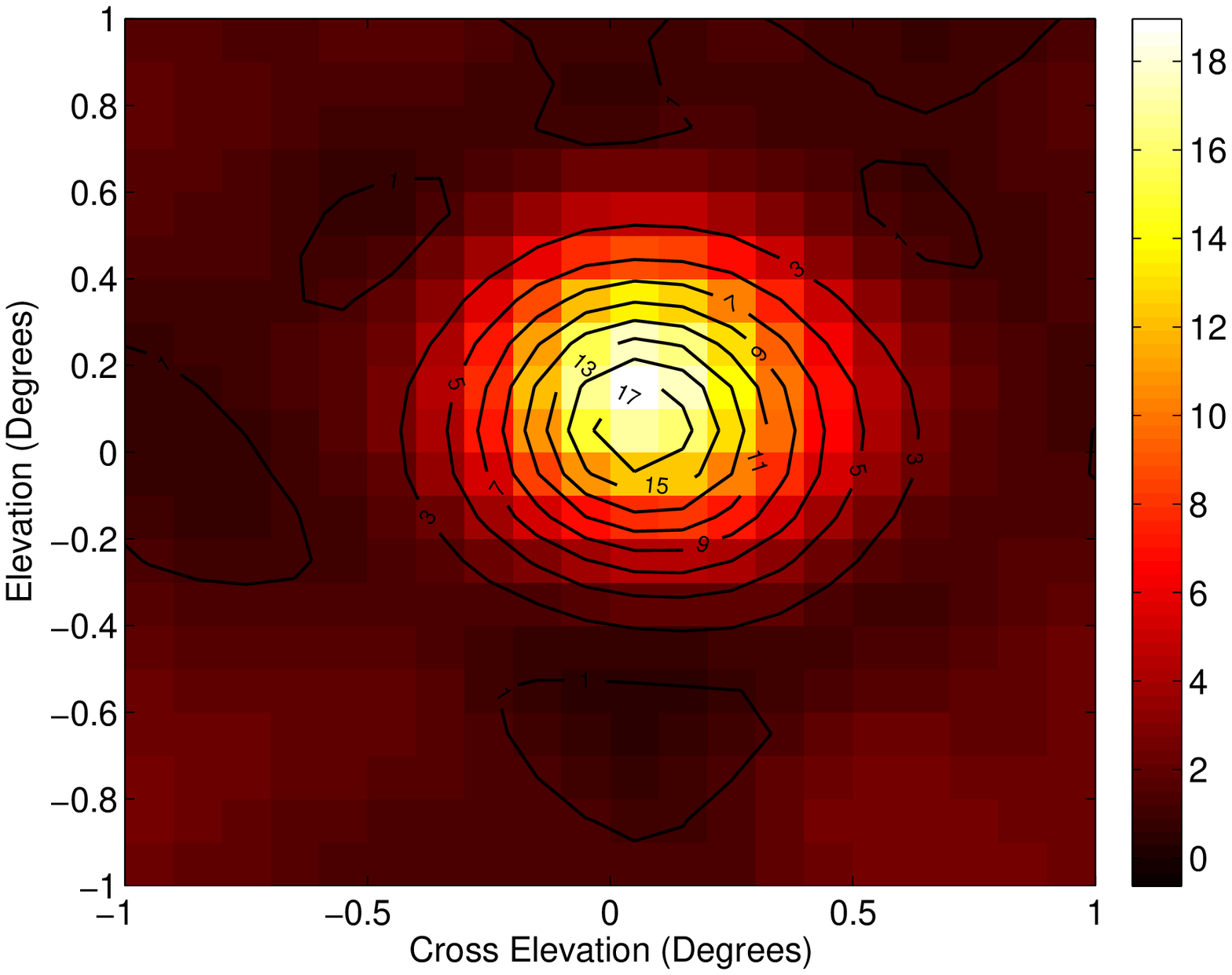}
\label{fig:3C295}
}
\centering
\subfigure[Mosaic image of the W49N region.]{
\includegraphics[width=3.0in]{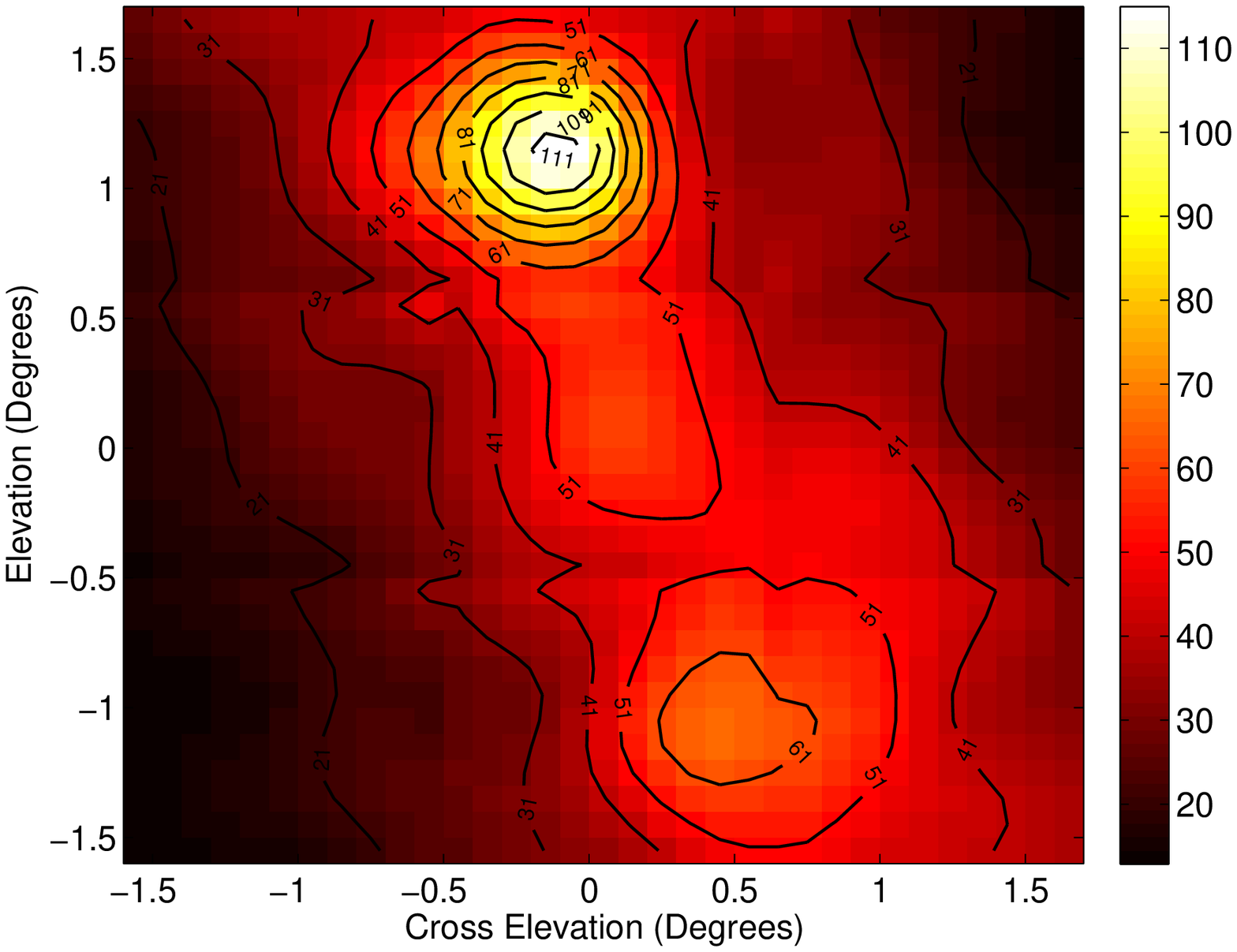}
\label{fig:W49N}
}
\caption{Radio camera image examples.  (a):  Single reflector pointing,
20 $\times$ 20 pixels, 400 total simultaneous beams.
(b):  3 $\times$ 3 mosaic of PAF images obtained with nine reflector pointings.  Each of the nine mosiac tiles has 11 $\times$ 11 pixels from 121 simultaneous beams.  Lower right is 3C397, and upper left is 3C398 (W49N).  All images are in units of Jy.}
\label{fig:radioCam}
\end{figure}

Figure \ref{fig:radioCam} presents examples of a single pointing image and an image mosaic.  The observation is over a 450 kHz band centered on the 1612 MHz OH line.  The continuum source 3C295 seen in Fig.~\ref{fig:3C295} has a flux density of 21 Jy at 1400 MHz.  The integration time was 60 seconds.
The radio camera grid is 20 $\times$ 20 pixels, consisting of a total of 400 simultaneously formed and electronically steered beams.  The 3 $\times$ 3 mosaic of Fig.~\ref{fig:W49N} shows OH source W49N and the nearby 3C397.  Both images were oversampled with more beams than necessary, but they illustrate the fine-scale radio imaging possible.  The number of pixels (beams) that can be formed from a data set is limited only by the time required to collect calibrations and the available computational capacity, since additional beams in post-correlation processing do not require more array elements or data samples.
A practical radio camera would likely form beams separated by on the order of half of the HPBW crossover distance.

\begin{figure*}[!t]
\hspace{.2in}
{
{\includegraphics[width=2.75in]{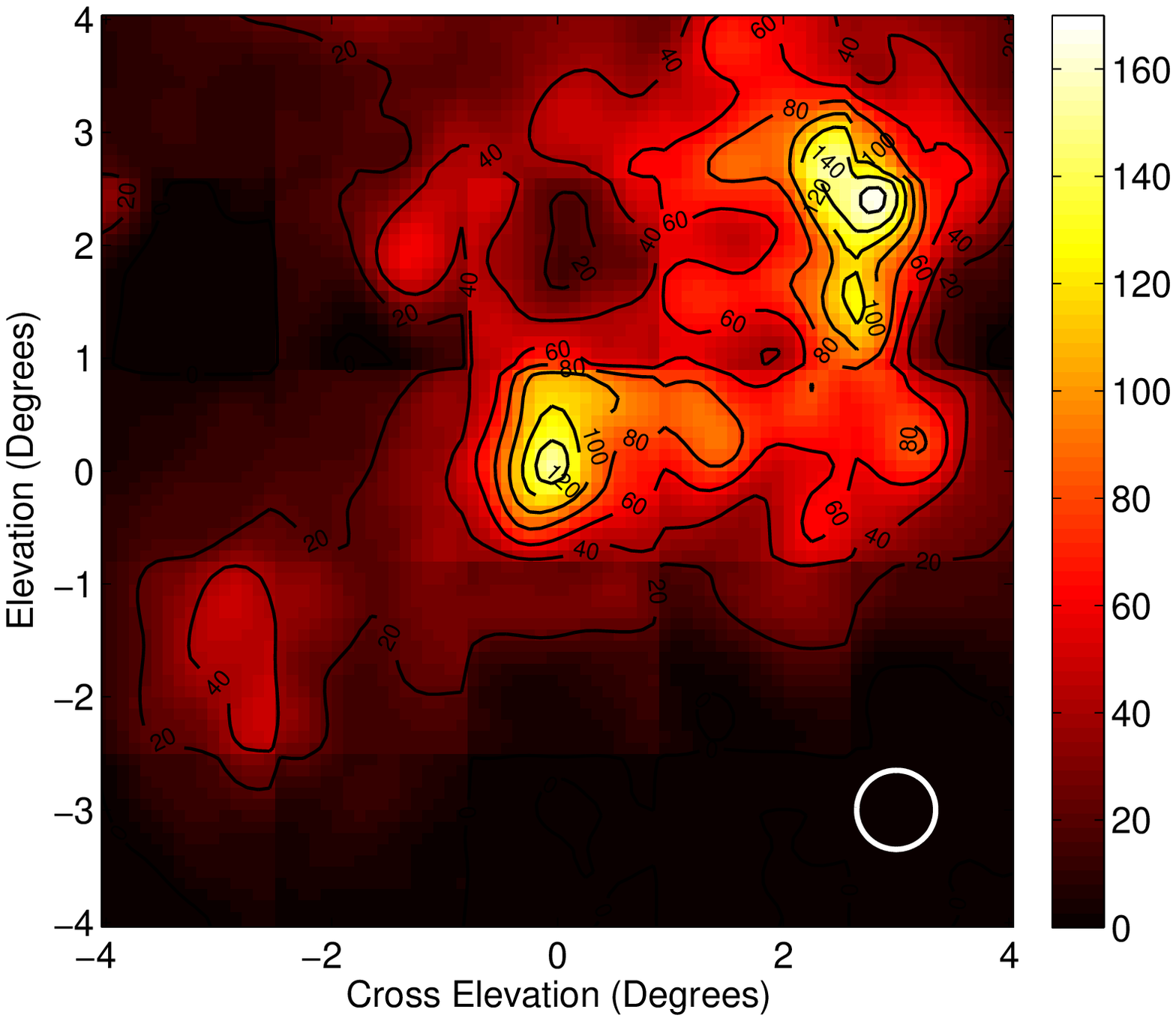}}}%
\hspace{.2in}
{
{\raisebox{.2in}{\includegraphics[width=3.2in]{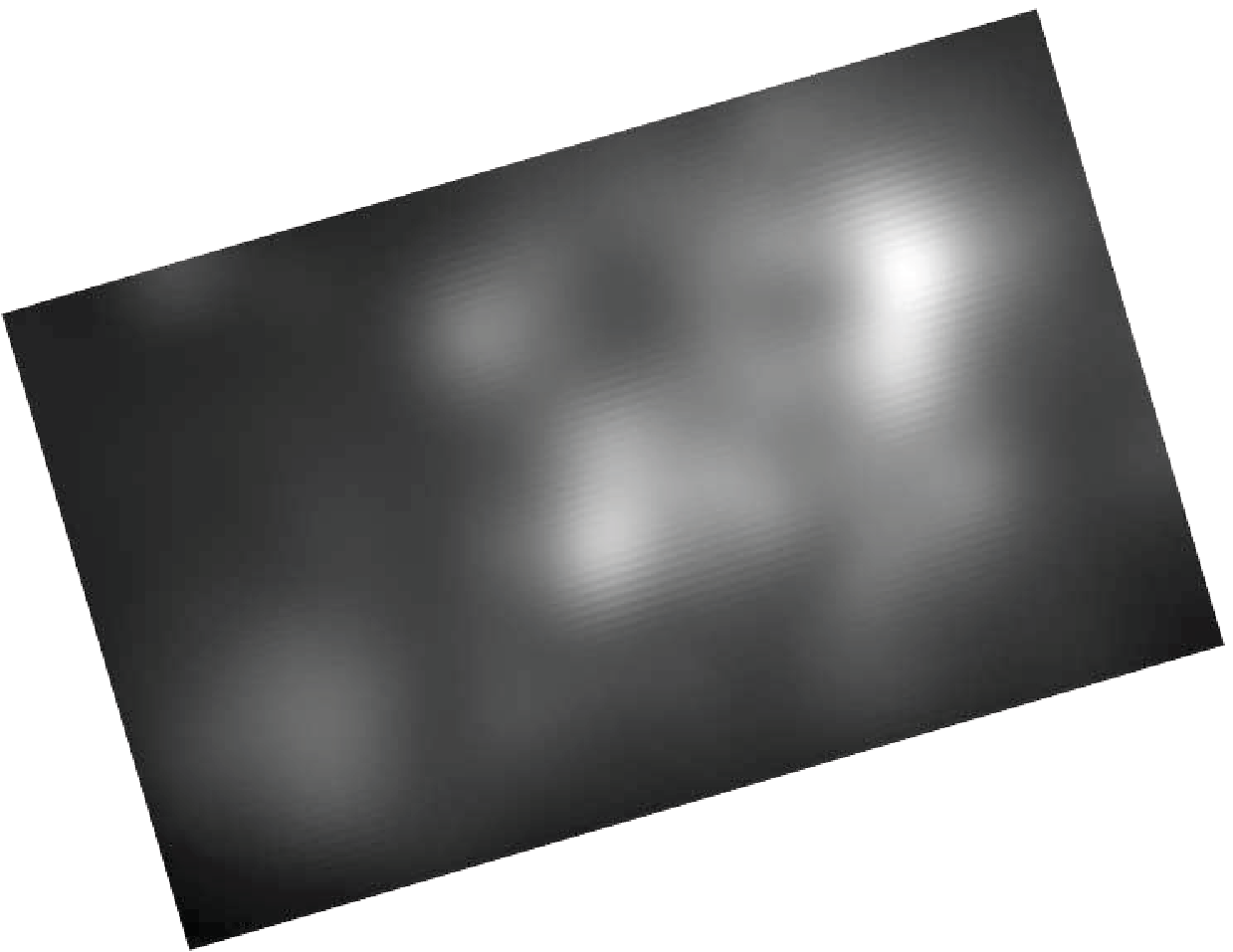}}}}
\caption{Left:  Cygnus X region at 1600 MHz.  5 $\times$ 5 mosaic of images using the 19-element prototype PAF on the Green Bank 20-Meter Telescope.  The circle indicates the half-power beamwidth.  Right:  Canadian Galactic Plane Survey image \citep{cgps_taylor_03} convolved to the 20-Meter effective beamwidth.  The center of the map is approximately 20$^{\rm h}$44$^{\rm m}$, +42$^{\rm o}$ (J2000) with north to the upper left.}
\label{fig:CygX}
\end{figure*}

Fig.~\ref{fig:CygX} presents a mosaic image of a more complex source distribution, the Cygnus X region, observed at 1600 MHz with a 5 $\times$ 5 mosaic for 25 total reflector pointings.  The inset circle indicates the HPBW beamwidth for a single pixel beam.  As a comparison, the right image is from the Canadian Galactic Plane Survey image, but blurred by convolution with the equivalent beam pattern of the 20$\,$m telescope to match resolution scales.  We expect that the image artifacts caused by discontinuities at mosaic tile boundaries could be eliminated with more sophisticated processing.

The Cygnus X radio camera image contains approximately 3000 pixels.
A more practical coarse grid spacing of about half the HPBW would require about 600 pixels.
A single horn feed would require 600 pointings (one for each pixel) to form such an image, compared to 25 (one for each mosaic tile) for the radio camera.
Thus for equal integration times per pixel, this radio camera provides an imaging speed up of 24 times.

\section{Elevation Dependent Sensitivity Optimization}
\label{sec:tipping}

PAFs introduce the possibility of adapting formed beam patterns to maximize sensitivity as the noise environment changes.  Since periodic noise measurements are required to form images, the information required to optimize beamformer weights periodically is already available as a byproduct of the observation. As the reflector pointing angle moves away from zenith, from the perspective of the feed the sky and spillover brightness temperature distributions change. In this section, we study the performance benefits of adaptation of the beam patterns to optimize sensitivity to account for changes in the sky and spillover noise distributions.


\subsection{Simulation Results}
\label{sec:simresults}

%

\begin{figure}[!t]
\centering
\subfigure[Mutually coupled LNA ($T_{\rm rec} $) and spillover  ($T_{\rm sp}$) noise temperatures.]{\includegraphics[width = 7.0cm]{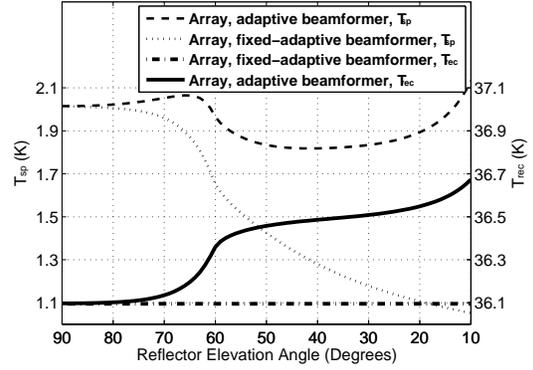}
    \label{fig:Tlna_sim}
    \label{fig:Tsp_sim}}
\subfigure[System noise temperature, $T_{\rm sys}$.]{\includegraphics[width = 7.0cm]{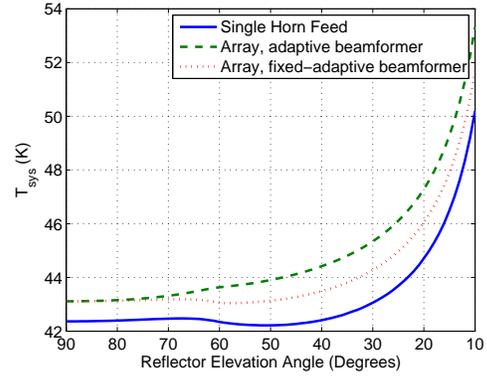}
    \label{fig:Tsys_sim}}
\caption{Noise temperature variations as a function of dish elevation. Active impedance matching is assumed for the array.}
\label{fig:sim_temps}
\end{figure}

In order to analyze the separate effects of constituent noise terms in $T_{\rm sys}$, the approach is to simulate expressions for $\mR_{\rm rec}$, $\mR_{\rm sp}$ and $\mR_{\rm sky}$ at each elevation angle, and then compute the corresponding noise temperatures as the beamformer weights change.
In order to provide an upper bound on achievable PAF performance metrics, where simulations are compared to a horn feed, the model incorporates active impedance matching (see Sec.~\ref{sec:hotCold}) at the LNA-antenna interface to minimize effects of coupling between array elements. This lowers the mutual coupling noise component of $T_{\rm rec}$ from the $23\,$K modeled value in Table \ref{tab:noise} to 3--4$\,$K (since active impedances depend on the beam steering direction, the LNAs cannot be perfectly matched to the array for all beam steering directions, and the mutual coupling component of the receiver noise varies over the PAF field of view).
The difficulty with achieving these improvements is that active impedances depends on $\vw$ and therefore requires a different matching network for each steered beam.  We are currently developing an array based on the methods of \cite{mc_warnick_08} which provides the best average match for all beams within the radio camera field of view and for all frequencies within the useful system bandwidth.  Preliminary results suggest that this approach realizes most of the available improvement in $T_{\rm sys}$ and sensitivity.
 The amplifier minimum noise temperature parameter $T_{\rm min}$ was $33\,$K and a constant $T_{\rm cmb} + T_{\rm gb} = 3\,$K was assumed at all elevations.

The minimum variance distortionless response (MVDR) beamformer described in Section \ref{sec:Beamformers} was used to compute a new $\vw^{(j)}$ at each elevation to adapt to changes in $\mR_{\rm n}$.  For this scenario, MVDR is essentially identical to the maximum sensitivity beamformer.  For comparison, results are also given for a fixed beamformer with elevation-independent weights obtained from the MVDR solution computed for the zenith pointing.

For the fixed beamformer curves in Fig.~\ref{fig:sim_temps}, there is no fluctuation in $T_{\rm rec}$ because both the beamformer weights and the input receiver noise power are independent of the pointing direction.  A decrease in $T_{\rm sp}$ occurs as more of the pattern sidelobes in the spillover region are directed away from warm ground toward a cooler sky.  $T_{\rm atm}$ (not shown) varies according to the secant curve described in Section~\ref{sec:main_beam}, increasing as the dish elevation angle approaches the horizon and is identical for both the fully adaptive and fixed--adaptive beamformers.  At Zenith $T_{\rm atm} = 2K$ while at an elevation of $10^\circ$ it rises to about 11K.


\begin{figure}[!t]
\centering
\subfigure[Aperture efficiency.]{\includegraphics[width = 7cm]{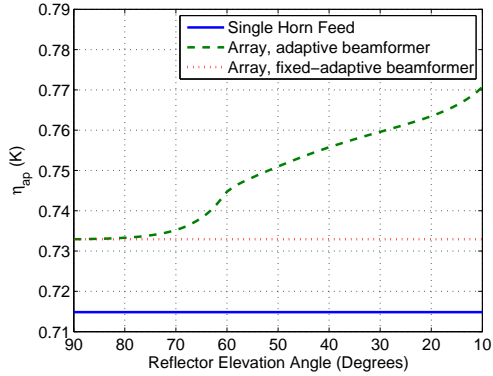}
   \label{fig:etaAp_sim}}
\subfigure[Sensitivity.]{\includegraphics[width = 7cm]{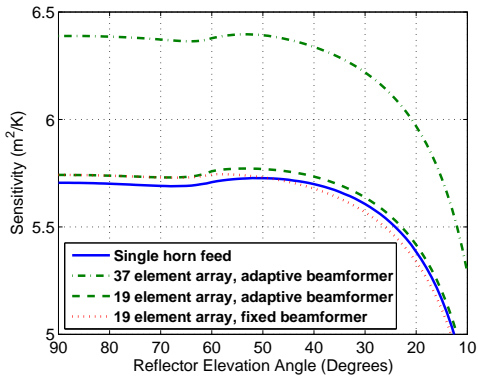}
    \label{fig:sens_sim}}
\caption{The 19 element adaptive beamformer provides an improvement in aperture efficiency and sensitivity as it responds to changes in the noise field. Higher sensitivity is achieved at
mid elevations than for either the fixed PAF beamformer or horn feed.  A 37 element PAF would further increase sensitivity. Active impedance matching is assumed for each array.}
\label{fig:ap_ef_sens}
\end{figure}

For the adaptive beamformer, results are less intuitive, since the system noise increases as the tipping angle moves away from zenith, as seen in Fig.~\ref{fig:Tsys_sim}.  As the reflector boresight moves closer to the horizon, the dominant change in the noise environment is the increase in the sky noise temperature in the antenna main beam.  Since the beamformer optimizes sensitivity, the increase in sky noise temperature leads to an increase in the beam aperture efficiency.  At zenith, the optimal beam corresponds to a reflector illumination pattern with a higher taper at the reflector edge, lower aperture efficiency, and higher spillover efficiency.  Near the horizon, increased sky noise means that the optimal beam corresponds to an illumination pattern with a lower taper and higher aperture efficiency.  This effect is apparent in Fig.~\ref{fig:etaAp_sim}.

Figures \ref{fig:Tsys_sim}--\ref{fig:sens_sim} compare PAF results for a single horn feed system.  The horn is modeled using the standard $\cos^q(\phi)$ illumination pattern model, where $\phi$ is the angle, from the perspective of the feed, along the dish surface relative to the boresight axis.  The value of $q$ controls illumination taper and aperture efficiency, and is chosen to maximize sensitivity at zenith.  This is a realistic approximate model for a typical single-pixel feed.  For a 20$\,$m reflector with $f/D = 0.43$, optimal sensitivity is obtained with $q = 4.6$, which produces a $-14$ dB illumination taper at the dish edge.  This comparison shows that the 19 element array, steered to boresight, with adaptive beamforming is capable of achieving a sensitivity at least as high as the typical single horn feed system.  As noted above, to make a fair comparison to a single feed, for this simulation the LNAs are active impedance matched to the array to reduce the mutual coupling noise penalty.

Figure \ref{fig:sens_sim} also includes a simulation for a 37 element array to show sensitivity improvements that are possible with a larger feed.  The 37 element PAF uses the same element design and inter-element spacing as the 19 element array.  The significantly higher sensitivity compared with both the standard horn feed and the 19 element array is due to its larger aperture and the ability to finely control current phase and amplitude distribution across the array to achieve a more nearly optimal illumination pattern.
This would be difficult to accomplish with standard horn feed technology, even with an equivalent aperture size.  The larger array also permits beams to be steered off-axis with a smaller sensitivity penalty \citep{Waldron08}.

\subsection{Experimental Results}
\label{sec:tipping_experimental}

\begin{figure}[!t]
\centering
\subfigure[System noise temperature, $T_{\rm sys}$.]{\includegraphics[width = 7cm]{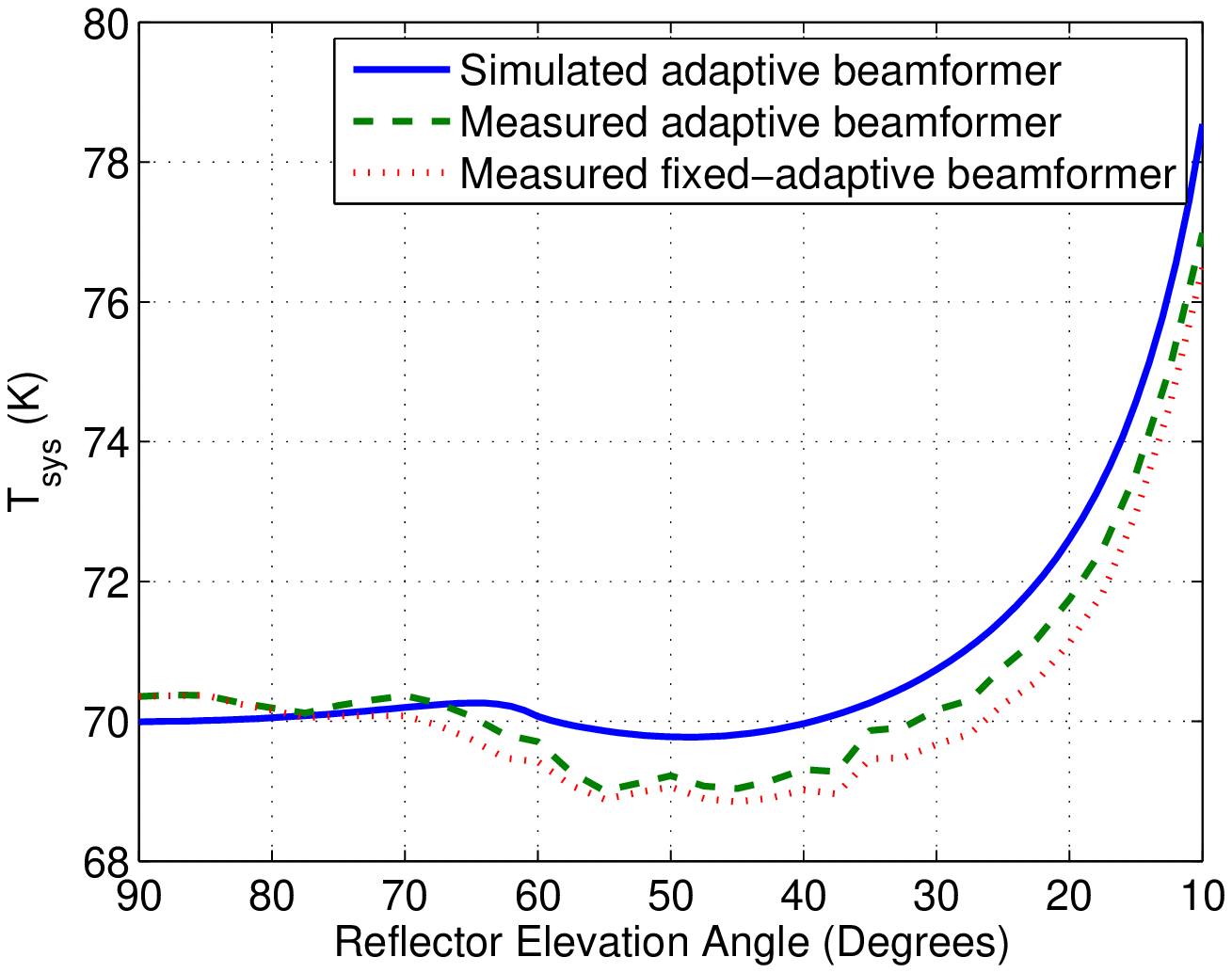}
   \label{fig:Tsys_sim_data}}
\subfigure[Sensitivity.]{\includegraphics[width = 7cm]{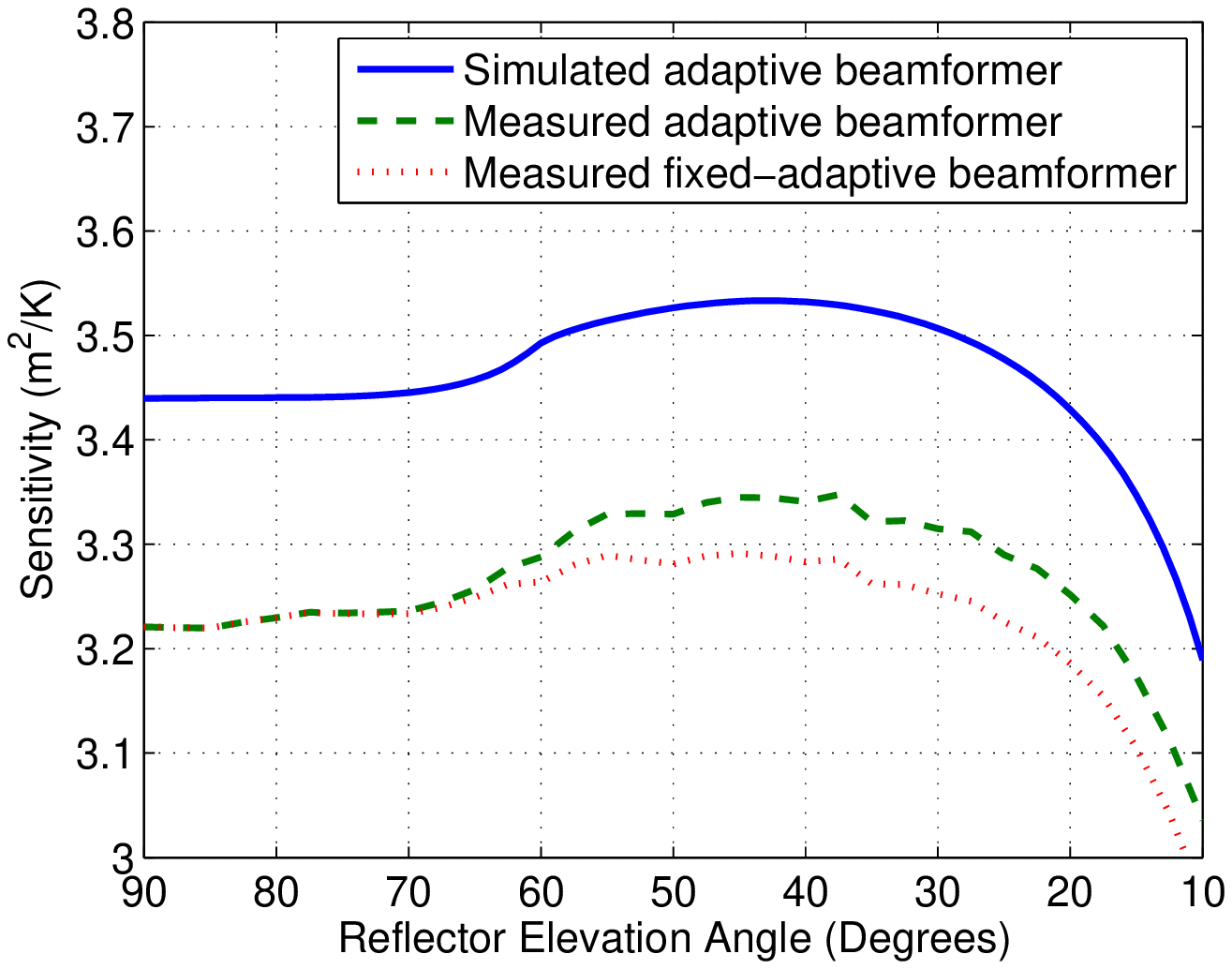}
    \label{fig:sens_sim_data}}
\caption{Real data results for 19 element adaptive beamforming while tipping the Green Bank 20-Telescope from zenith to the horizon.  Simulated sensitivity is higher primarily due to scattering and blockage effects neglected in the model.  The close curve shape match between simulated and measured results validates the simulation model.  The adaptive beamformer exhibits improved sensitivity in the mid elevations.  The physical array and simulation both used element-wise self impedance matching.
 }
\label{fig:realTilt}
\end{figure}





The simulated dish tipping scenario of the previous section was applied to a real data acquisition experiment with the 19 element PAF test platform on the 20m dish at the NRAO observatory in Green Bank, West Virginia.
The following results use data with no bright sources in the array field of view collected during the July 2008 experiment.  The observation band was centered at $1612$ MHz with a bandwidth of $450$ kHz.  Array data was recorded for telescope pointings ranging from zenith to the horizon in 2.5--degree increments.

MVDR beamformer weight updates were implemented as described in Section \ref{sec:Beamformers} using $\hat{\mR}^{(j)}$ to compute a new $\vw^{(j)}$ at each elevation for the adaptive beamformer.
Calculations for sensitivity, aperture efficiency, and system temperature were formed using elevation independent estimates of $\breve{\mR}_{{\rm s},i}$ and $\breve{\mR}_{\rm n}$ from calibration data for the boresight beam, and the elevation dependent $\vw^{(j)}$ in (\ref{eq:sensitivSNR}), \eqref{eq:eta_ap} and (\ref{eq:Tsys}).

Experimental results for $T_{\rm sys}$ and sensitivity are shown in Fig.~\ref{fig:realTilt}, along with simulations for comparison.  Simulated curves are based on the full noise model described in Section \ref{sec:noise_model}.
Since our active matched array is still under development, the 19 element test platform array and corresponding simulations reported in this section used element-wise self impedance matching (i.e., the LNAs are noise matched to the element self impedances, not the active impedances).  Comparison of Figs.~\ref{fig:Tsys_sim_data} and \ref{fig:Tsys_sim} illustrates that active impedance matching (see Sec.\ \ref{sec:hotCold}) has the potential for this array of reducing $T_{\rm sys}$ by 23$\,$K.  Ohmic losses in Fig.\ \ref{fig:Tsys_sim} are assumed to be negligible, whereas Fig.\ \ref{fig:Tsys_sim_data} includes 4$\,$K noise due to loss. Figures \ref{fig:sens_sim_data} and \ref{fig:sens_sim} show that active impedance matching can increase sensitivity by a factor of 1.67 (2.2 dB).

There is significant agreement in Fig.~\ref{fig:realTilt} between experimental real data and simulated results, which serves as a validation for the proposed noise models.  In particular, the increase in $T_{\rm sys}$, $\eta_{\rm ap}$, and the modest improvement in sensitivity at lower elevations for the adaptive beamformer with respect to the fixed beamformer predicted by the model simulations has been verified with real data.
Computed real data aperture efficiency (not shown) also exhibited an excellent match with the model simulations.
The good match in this example application suggests that the model may be used with confidence in a variety of other signal scenarios.

In some observing scenarios it may not be practical to obtain an SOI--free $\hat{\mR}^{(j)}$ estimate to update $\vw^{(j)}$ continuously  with elevation changes.  In our example we took pains to find a relatively source-free patch of sky for the elevation scan.  Since elevation dependence of sky and spillover noise is based primarily on the geometry and not on transient sources, however, it is possible to pre-compute a set of optimal, elevation indexed beamformers from calibration data.  These could then be called up at a later time (over several days) from a look-up table to achieve the improved mid-elevation sensitivity shown above.

The degradation rate of calibration data due to phase and amplitude drift in the receiver electronics and other effects is currently under study, as are techniques to refresh calibration data using a small number of bright source observations or external electronic calibration sources.


\section{RFI Mitigation}

\label{sec:rfi}

As contemporary science goals increasingly require
observing sources outside the traditional protected spectrum bands, a
critical need is developing to deal with ubiquitous man-made
interfering signals such as satellite downlink transmissions
\citep{Ellingson01,Poulsen05,Combrinck94}, radar systems
\citep{Dong05,Jeffs06,Fisher01b,Ellingson03b,Zhang03}, air navigation
aids \citep{Zhang05,Fisher05}, wireless communications
\citep{Leshem99b,vdVeen2000}, and digital television broadcasts.
Even locating instruments in undeveloped areas with regulatory protection
does not avoid many man-made sources such as satellite downlinks.

The PAF, as illustrated in Fig.~\ref{fig:multibeam}, offers a
promising new approach that exploits the {\em spatial} structure of the interfering signal to track and remove it without having to discard data \citep{Jeffs08b}.
With spatial canceling, even interferers that entirely overlap the signal of
interest (SOI) spectrum may be mitigated.
It is anticipated that using this technique, data collection may now be possible where
previously interference was too dominant to permit serious scientific observations.
\begin{figure}[!t]
\begin{center}
\subfigure[No RFI.]{
\includegraphics[width=1.7in]{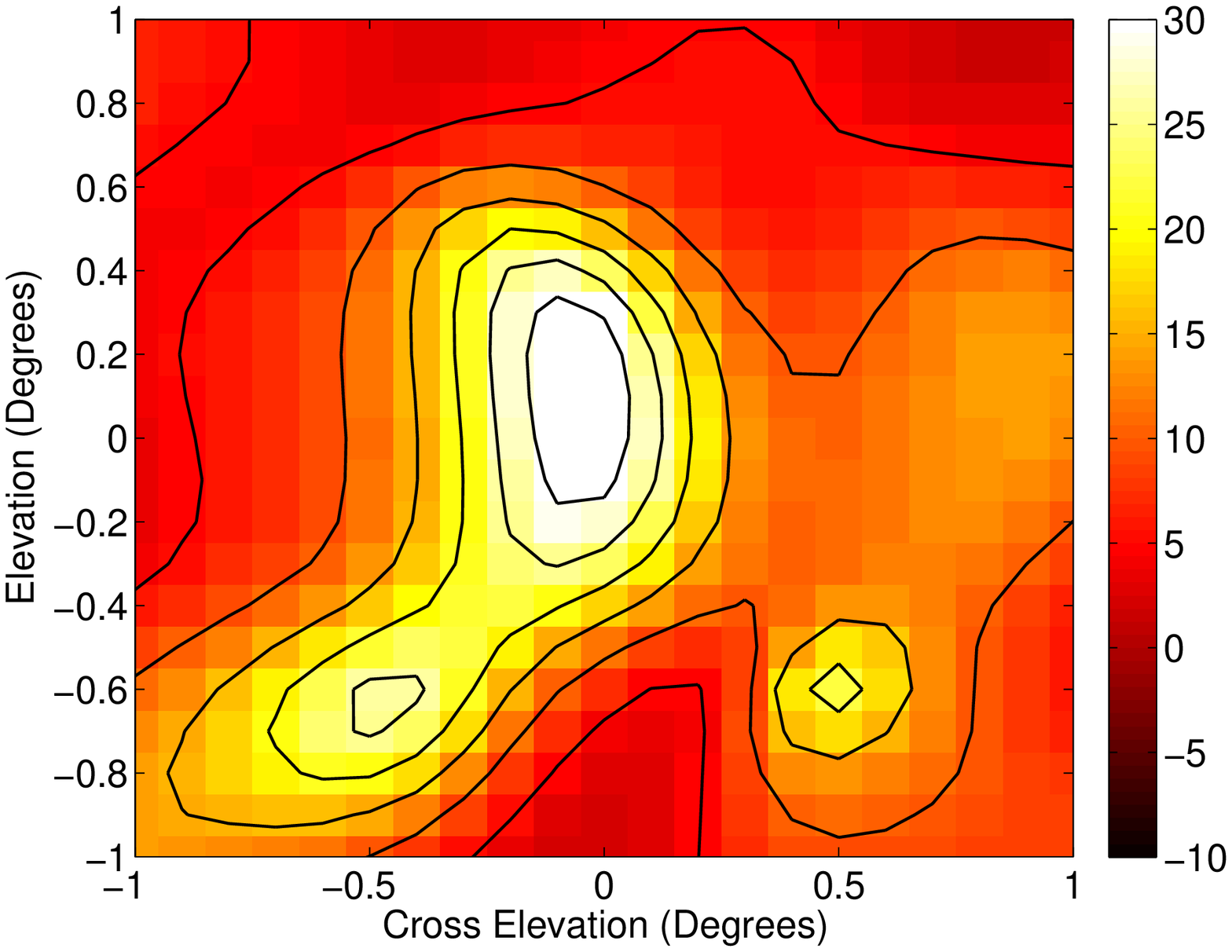}
}
\subfigure[With RFI.]{
\includegraphics[width=1.7in]{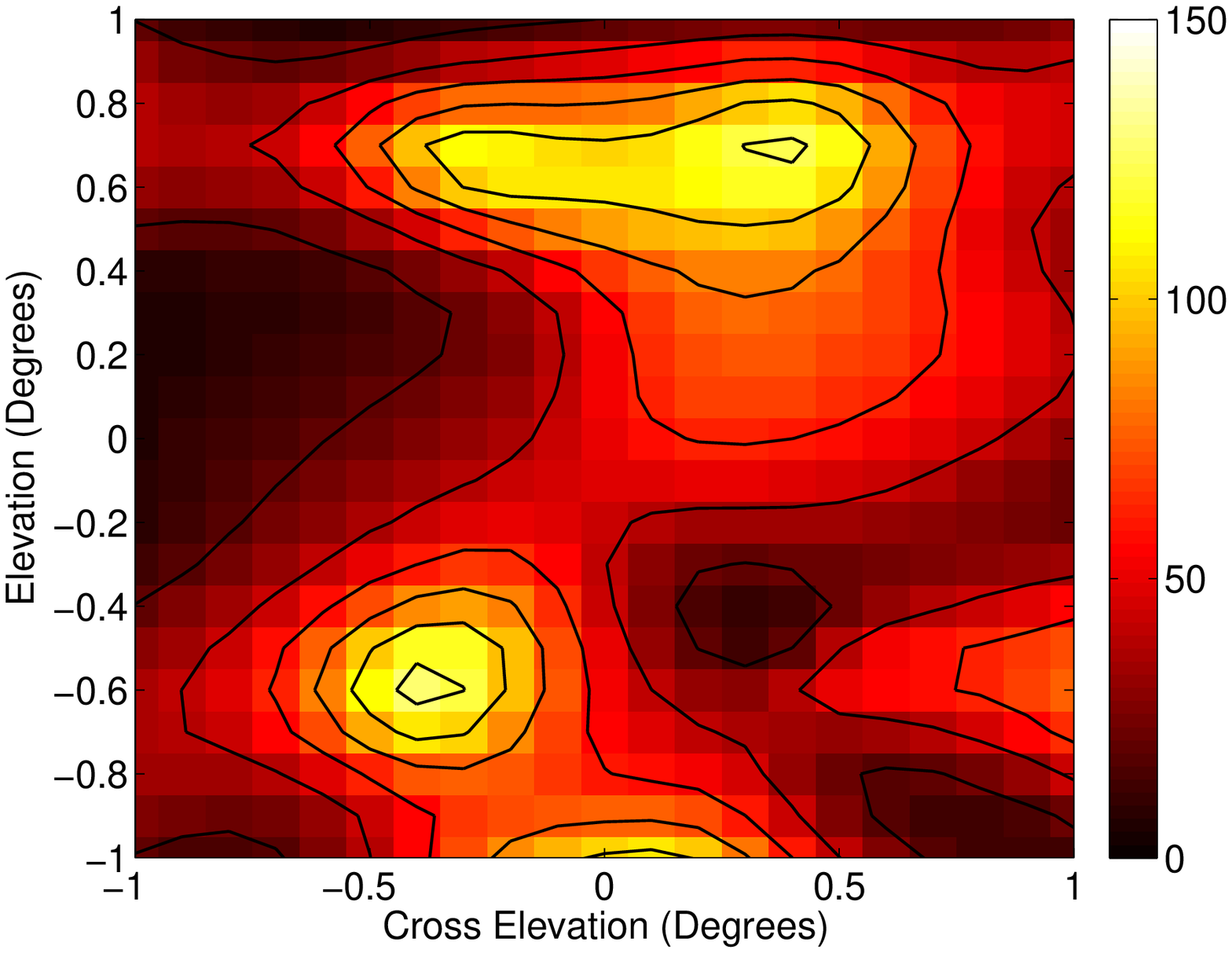}
}
\subfigure[RFI canceled.]{
\includegraphics[width=1.7in]{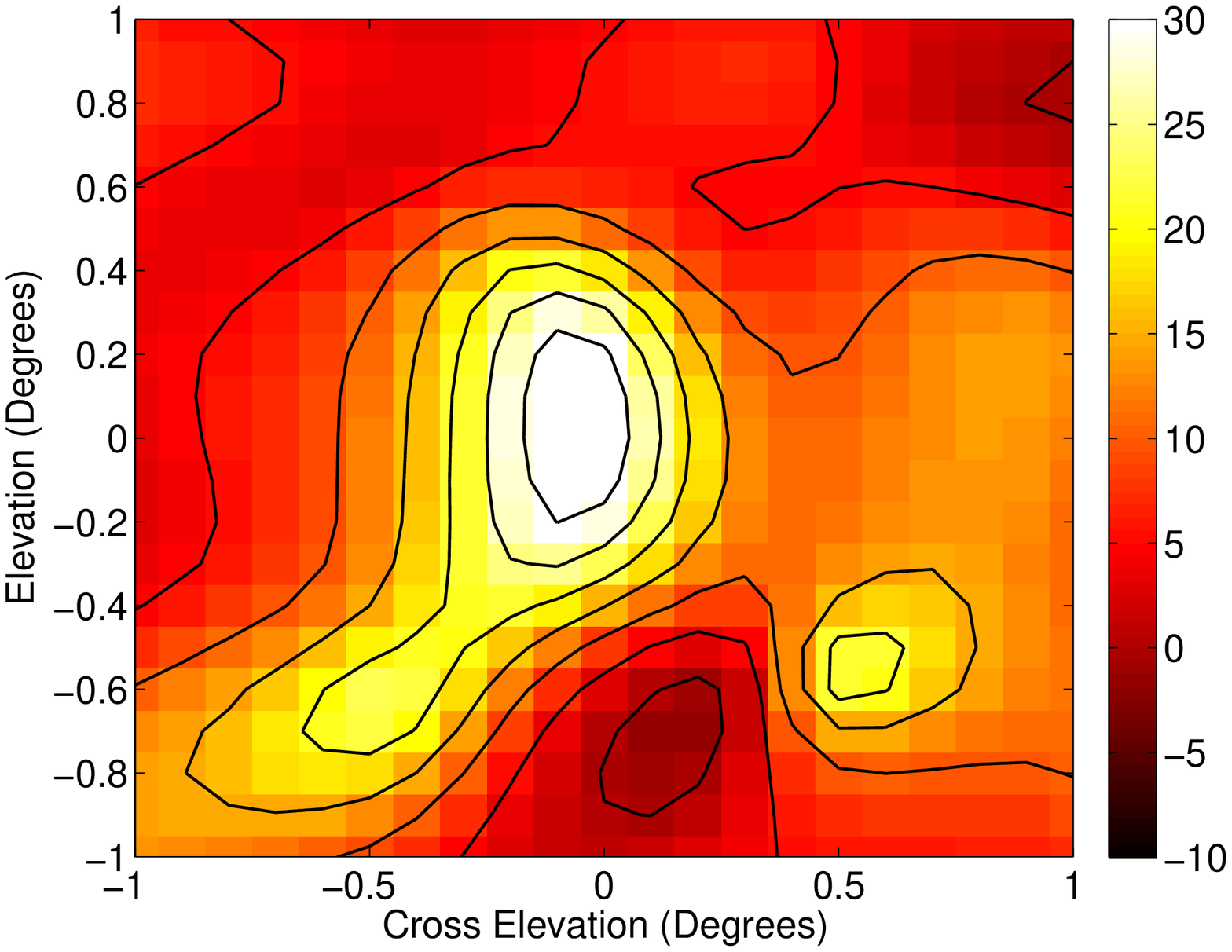}
}
\end{center}
\caption{W3OH image with and without RFI.  The color scale is equivalent antenna temperature (K). }
\label{fig:OHrfi}
\end{figure}

To demonstrate feasibility of adaptive RFI mitigation with the 19 element array on the 20m dish, a local, moving, man-made interference source was introduced.
This involved an RF function generator and hand-held antenna, while walking a circuit located about one kilometer from the telescope.  The interfering signal was FM-modulated with a 200 kHz bandwidth overlapping and masking the W3OH spectral line seen at the array at 1665 MHz.
RFI was removed using the subspace projection algorithm \citep{Leshem00,Jeffs08b}.
This method adapts the beamformer weights to the changing interference environment, so as to place a spatial null in the interference direction.
Beamformer weights are updated every STI according to
\begin{eqnarray}
\vw^{(j)}_{{\rm sp},i} & = & \mP^\perp \vw_i \nonumber \\
\mP^\perp & = & \mI - \mU_{\rm int} \mU_{\rm int}^H , \nonumber \\
\hat{\mR}^{(j)} [  \mU_{\rm int} \, | \,  \mU_{\rm n} ] & = & [  \mU_{\rm int} \, | \,  \mU_{\rm n} ] \,\boldsymbol{\Lambda}
\label{eqn:partition}
\end{eqnarray}
where $j$ is the STI index, sample covariance $\hat{\mR}^{(j)}$ is computed as in (\ref{eqn:R_x_hat}),
and $\vw_j$ is the nominal deterministic (non adaptive canceling) beamformer weight vector for the beam steered in the $i$th direction.
Equation (\ref{eqn:partition}) is an eigenvector decomposition of $\hat{\mR}^{(j)}$ with eigenvalues in diagonal $\boldsymbol{\Lambda}$ ordered in descending magnitude such that the dominant eigenvectors in
the partition $\mU_{\rm int}$ span the interference subspace.
The full eigenvector matrix $\mU = [  \mU_{\rm int} \, | \,  \mU_{\rm n} ]$ is unitary.

Images of the source with and without RFI mitigation are shown in Fig.~\ref{fig:OHrfi}. Some small distortion due to residual RFI is apparent in Fig.~\ref{fig:OHrfi}(c) after subspace projection adaptive cancelation processing, but the source which was completely obscured by interference is now clearly visible.


\section{Conclusions}

We have reported experimental results for a prototype L-band PAF and have demonstrated radio camera imaging using in situ calibrated maximum sensitivity beamforming.  Experimental results agree with simulations for a complete electrical and noise model for the PAF and reflector system.  Measured beam sensitivities, aperture efficiencies and system noise temperatures achieve expectations for the 19 element, uncooled array.  The potential of PAFs for adaptive RFI mitigation was explored.  These results indicate that there is a clear path forward to a high sensitivity, wide field of view PAF with scientifically useful performance.

Ongoing work includes experimental demonstration of active imped\-ance matching to reduce the noise contribution caused by array mutual coupling, studies of beamformer design procedures with controlled pattern shape, development of an L-band PAF and associated back-end signal processing for the Green Bank Telescope, and cryogenic arrays that achieve system temperatures of approximately 20$\,$K.


\acknowledgments

This work was funded by National Science Foundation under grant number AST 0352705.  We express appreciation to Jay Lockman for providing examples of the science that will be enabled by phased array feeds.

\section*{Appendix}

The following notations are used in the paper:
\begin{enumerate}

\item $E\{ \mA \}$ : Expected value of random $\mA$.

\item $\bar{E}_m(\Omega)$:  Far-field electric field pattern at spherical angle $\Omega$ due to array element m.

\item $\lfloor a \rfloor$ : Floor operation, rounding toward zero.

\item $a^*$ : Complex conjugate of $a$.

\item $\mA^T$, $\mA^H$ : Transpose and conjugate transpose of $\mA$.

\item $\hat{\mA}$ : Sample estimate of some parameter, $\mA$.

\item $\breve{\mA}$ : Calibration data set estimate of $\mA$.

\item ${\mI}$ : The identity matrix.

\end{enumerate}

\bibliographystyle{apj}


\begin{thebibliography}{54}
\expandafter\ifx\csname natexlab\endcsname\relax\def\natexlab#1{#1}\fi

\bibitem[{{A. R. Taylor et al.}(2003)}]{cgps_taylor_03}
{A. R. Taylor et al.} 2003, Astronomical Journal, 125, 3145

\bibitem[{Black \& Dalgarno(1977)}]{black_77}
Black, J.~H., \& Dalgarno, A. 1977, ApJS, 34, 405

\bibitem[{Boonstra {et~al.}(2000)Boonstra, Leshem, {van der Veen}, Kokkeler, \&
  Schoonderbeek}]{vdVeen2000}
Boonstra, A., Leshem, A., {van der Veen}, A.-J., Kokkeler, A., \&
  Schoonderbeek, G. 2000, in Proc. of the IEEE International Conf. on Acoust.,
  Speech, and Signal Processing, Vol.~6, 3546--3549

\bibitem[{Braun \& Thilker(2004)}]{braun_04}
Braun, R., \& Thilker, D.~A. 2004, A\&A, 417, 421

\bibitem[{Combrinck {et~al.}(1994)Combrinck, West, \& Gaylard}]{Combrinck94}
Combrinck, W., West, M., \& Gaylard, M. 1994, Astronomical Society of the
  Pacific, 106, 807

\bibitem[{{D. A. Thilker et al.}(2004)}]{thilker_04}
{D. A. Thilker et al.} 2004, Astrophys. J., 601, L39

\bibitem[{Dame \& Thaddeus(2008)}]{dame_08}
Dame, T.~M., \& Thaddeus, P. 2008, Astrophys. J., 683, L143

\bibitem[{Dong {et~al.}(2005)Dong, Jeffs, \& Fisher}]{Dong05}
Dong, W., Jeffs, B., \& Fisher, J. 2005, Radio Science, {\em RS5S04,
  doi:10.1029/2004RS003130}, 40

\bibitem[{{E. E. M. Woestenburg}(2005)}]{active_woestenburg_tr_05}
{E. E. M. Woestenburg}. 2005, Noise matching in dense phased arrays, Tech. Rep.
  RP-083, ASTRON, Dwingeloo, The Netherlands

\bibitem[{Ellingson \& Hampson(2003)}]{Ellingson03b}
Ellingson, S., \& Hampson, G. 2003, The Astrophysical Journal Suplement Series,
  147, 167

\bibitem[{et~al.(2008{\natexlab{a}})}]{chynoweth_08}
et~al., K. M.~C. 2008{\natexlab{a}}, Astronomical J., 135, 1983

\bibitem[{et~al.(2008{\natexlab{b}})}]{leach_08}
et~al., S. M.~L. 2008{\natexlab{b}}, A\&A, 491, 597

\bibitem[{Fisher(2001)}]{Fisher01b}
Fisher, J. 2001, Analysis of Radar Data from February 6, 2001, Tech. rep.,
  National Radio Astronomy Observatory, Green Bank Observatory

\bibitem[{Fisher {et~al.}(2005)Fisher, Zhang, Y.~Zheng, \& Bradley}]{Fisher05}
Fisher, J., Zhang, Q., Y.~Zheng, S.~W., \& Bradley, R. 2005, The Astronomical
  Journal, 129, 2940

\bibitem[{Gilmon \& Shull(2006)}]{gillmon_06}
Gilmon, K., \& Shull, J.~M. 2006, Astrophys. J., 636, 908

\bibitem[{Ivashina {et~al.}(2008)Ivashina, Maaskant, \&
  Woestenburg}]{array_ivashina_08}
Ivashina, M.~V., Maaskant, R., \& Woestenburg, B. 2008, IEEE Ant. Wireless
  Propag. Lett., 7, 733

\bibitem[{Jeffs {et~al.}(2006)Jeffs, Lazarte, \& Fisher}]{Jeffs06}
Jeffs, B., Lazarte, W., \& Fisher, J. 2006, Radio Science, 41,
  doi:10.1029/2005RS003400

\bibitem[{Jeffs \& Warnick(2007)}]{Jeffs07icassp}
Jeffs, B., \& Warnick, K. 2007, in Proc. of the IEEE International Conf. on
  Acoust., Speech, and Signal Processing, Vol.~II, Honolulu, 1145--1148

\bibitem[{Jeffs \& Warnick(2008)}]{Jeffs07}
Jeffs, B., \& Warnick, K. 2008, {IEEE} Trans. Signal Processing, 56, 3108

\bibitem[{Jeffs {et~al.}(2008{\natexlab{a}})Jeffs, Warnick, Elmer, Landon,
  Waldron, Jones, Fisher, \& Norrod}]{Jeffs08c}
Jeffs, B., Warnick, K., Elmer, M., Landon, J., Waldron, J., Jones, D., Fisher,
  R., \& Norrod, R. 2008{\natexlab{a}}, in Calibration and Imaging Workshop,
  CALIM2008, Perth, Australia

\bibitem[{Jeffs {et~al.}(2009)Jeffs, Warnick, Landon, Elmer, Fisher, \&
  Norrod}]{Jeffs09}
Jeffs, B., Warnick, K., Landon, J., Elmer, M., Fisher, J., \& Norrod, R. 2009,
  in Calibration and Imaging Workshop, CALIM2009, Socorro, New Mexico

\bibitem[{Jeffs {et~al.}(2008{\natexlab{b}})Jeffs, Warnick, Landon, Waldron,
  D.~Jones, \& Norrod}]{Jeffs08b}
Jeffs, B., Warnick, K., Landon, J., Waldron, J., D.~Jones, J.~F., \& Norrod, R.
  2008{\natexlab{b}}, IEEE Journal of Selected Topics in Signal Processing, 2,
  635

\bibitem[{Landon {et~al.}(2008{\natexlab{a}})Landon, Jeffs, Warnick, Fisher, \&
  Norrod}]{Landon08}
Landon, J., Jeffs, B., Warnick, K., Fisher, J., \& Norrod, R.
  2008{\natexlab{a}}, in Proceedings of the URSI General Assembly, Chicago

\bibitem[{Landon {et~al.}(2008{\natexlab{b}})Landon, Jones, Jeffs, Wanick,
  Fisher, \& Norrod}]{Landon08b}
Landon, J., Jones, D., Jeffs, B., Wanick, K., Fisher, R., \& Norrod, R.
  2008{\natexlab{b}}, in Proceedings of the URSI National Radio Science
  Meeting, Boulder, CO

\bibitem[{Leshem {et~al.}(2000)Leshem, {van der Veen}, \& Boonstra}]{Leshem00}
Leshem, A., {van der Veen}, A.-J., \& Boonstra, A.-J. 2000, Astrophysical
  Journal Supplements, 131, 355

\bibitem[{Leshem {et~al.}(1999)Leshem, {van der Veen}, \&
  Deprettere}]{Leshem99b}
Leshem, A., {van der Veen}, A.-J., \& Deprettere, E. 1999, in Second IEEE
  Workshop on Signal Processing Advances in Wireless Communications, 1999,
  374--377

\bibitem[{Liszt \& Lucas(1996)}]{liszt_96}
Liszt, H., \& Lucas, R. 1996, A\&A, 314, 917

\bibitem[{Lockman {et~al.}(2008)Lockman, Benjamin, Heroux, \&
  Livingston}]{lockman_08}
Lockman, F.~J., Benjamin, R.~A., Heroux, A.~J., \& Livingston, G.~I. 2008,
  Astrophys. J., 679, L21

\bibitem[{{M. Grossi et al.}(2008)}]{grossi_08}
{M. Grossi et al.} 2008, A\&A, 487, 161

\bibitem[{Murphy(1989)}]{thesis_murphy}
Murphy, E.~M. 1989, Master's thesis, University of Virginia

\bibitem[{Nagel {et~al.}(2007)Nagel, Warnick, Jeffs, Fisher, \&
  Bradley}]{array_feed_exp_07}
Nagel, J.~R., Warnick, K.~F., Jeffs, B.~D., Fisher, J.~R., \& Bradley, R. 2007,
  Radio Science, 42, doi:10.1029/2007RS003630

\bibitem[{Oosterloo {et~al.}(2008)Oosterloo, van Cappellen, \&
  Bakker}]{Oosterloo08}
Oosterloo, T., van Cappellen, W., \& Bakker, L. 2008, in Calibration and
  Imaging Workshop, CALIM2008, Perth, Australia

\bibitem[{Poulsen {et~al.}(2005)Poulsen, Jeffs, Warnick, \& Fisher}]{Poulsen05}
Poulsen, A., Jeffs, B., Warnick, K., \& Fisher, J. 2005, Astronomical Journal,
  130, 2916

\bibitem[{Putman(2006)}]{putman_06}
Putman, M.~E. 2006, Astrophys. J., 645, 1164

\bibitem[{Roddy(2006)}]{roddy2006}
Roddy, D. 2006, Satellite Communications, 4th edn. (McGraw-Hill)

\bibitem[{{S. Borthakur et al.}(2008)}]{borthakur_08}
{S. Borthakur et al.} 2008, in AIP Conf. Proc., Vol. 1035, The Evolution of
  Galaxies through the Neutral Hydrogen Window, 201--204

\bibitem[{{S. G. Hay et al.}(2007)}]{askap_eucap_07}
{S. G. Hay et al.} 2007, in Proceedings of 2nd European Conference on Antennas
  and Propagation, Edinburgh, UK

\bibitem[{S.W.~Ellingson(2001)}]{Ellingson01}
S.W.~Ellingson, J.~Bunton, J.~B. 2001, Astrophysical J. Supplement, 135, 87

\bibitem[{{Van Trees}(2002)}]{VanTrees02IV}
{Van Trees}, H. 2002, Detection, Estimation, and Modulation Theory, Part IV,
  Optimum Array Processing (John Wiley and Sons)

\bibitem[{Van~Veen \& Buckley(1988)}]{vanveen1988}
Van~Veen, B., \& Buckley, K. 1988, ASSP Magazine, IEEE [see also IEEE Signal
  Processing Magazine], 5, 4

\bibitem[{Verheijen {et~al.}(2008)Verheijen, Osterloo, {van Cappellen}, Bakker,
  Ivashina, \& {van der Hulst}}]{apertif_08}
Verheijen, M. A.~W., Osterloo, T.~A., {van Cappellen}, W.~A., Bakker, L.,
  Ivashina, M.~V., \& {van der Hulst}, J.~M. 2008, arXiv:0806.0234v1

\bibitem[{Wakker \& van Woerden(1997)}]{wakker_97}
Wakker, B.~P., \& van Woerden, H. 1997, Ann Rev Astr Ap, 35, 217

\bibitem[{Waldron(2008)}]{Waldron08}
Waldron, J.~S. 2008, Master's thesis, Brigham Young University

\bibitem[{Warnick \& Jeffs(2006)}]{warnick2006}
Warnick, K., \& Jeffs, B. 2006, Antennas and Wireless Propagation Letters,
  IEEE, 5, 499

\bibitem[{Warnick \& Jensen(2005)}]{warnick2005}
Warnick, K., \& Jensen, M. 2005, Antennas and Propagation, IEEE Transactions
  on, 53, 2490

\bibitem[{Warnick {et~al.}(2007)Warnick, Waldron, Landon, Lilrose, \&
  Jeffs}]{Warnick07antProp}
Warnick, K., Waldron, J., Landon, J., Lilrose, M., \& Jeffs, B. 2007, in
  Proceedings of the 2nd European Conference on Antennas and Propagation,
  Edinburgh, UK

\bibitem[{Warnick \& Jeffs(2008)}]{effic_warnick_08}
Warnick, K.~F., \& Jeffs, B.~D. 2008, IEEE Antennas and Wireless Propagation
  Letters, 7, 565

\bibitem[{Warnick {et~al.}(2008)Warnick, Jeffs, Landon, Waldron, Fisher, \&
  Norrod}]{paf_warnick_ursi08}
Warnick, K.~F., Jeffs, B.~D., Landon, J., Waldron, J., Fisher, R., \& Norrod,
  R. 2008, in Proceedings of URSI General Assembly, Chicago, IL

\bibitem[{Warnick {et~al.}(2009{\natexlab{a}})Warnick, Jeffs, Landon, Waldron,
  Jones, Fisher, \& Norrod}]{PAF_iwat09_warnick}
Warnick, K.~F., Jeffs, B.~D., Landon, J., Waldron, J., Jones, D., Fisher,
  J.~R., \& Norrod, R. 2009{\natexlab{a}}, in Proceedings of the 2009 IEEE
  International Workshop on Antenna Technology

\bibitem[{Warnick \& Jensen(2007)}]{opt_match}
Warnick, K.~F., \& Jensen, M.~A. 2007, IEEE Transactions on Antennas and
  Propagation, 55, 1726

\bibitem[{Warnick {et~al.}(2009{\natexlab{b}})Warnick, Woestenburg,
  Belostotski, \& Russer}]{mc_warnick_08}
Warnick, K.~F., Woestenburg, B., Belostotski, L., \& Russer, P.
  2009{\natexlab{b}}, IEEE Trans. Ant. Propag., 57, 1634

\bibitem[{Willis(2009)}]{Willis09}
Willis, T. 2009, in Calibration and Imaging Workshop, CALIM2009, Socorro, New
  Mexico

\bibitem[{Zhang {et~al.}(2003)Zhang, Zheng, Wilson, Fisher, \&
  Bradley}]{Zhang03}
Zhang, Q., Zheng, Y., Wilson, S., Fisher, J., \& Bradley, R. 2003, The
  Astronomical Journal, 126, 1588

\bibitem[{Zhang {et~al.}(2005)Zhang, Zheng, Wilson, Fisher, \&
  Bradley}]{Zhang05}
---. 2005, The Astronomical Journal, 129, 2933

\end{thebibliography}

\end{document}